\documentclass[11pt,a4paper]{article}
\pdfoutput=1
\usepackage{amsmath,amssymb,color,epsfig,cite}
\usepackage{graphicx}
\usepackage{subfigure}
\usepackage{setspace}

\usepackage{amsfonts}

\allowdisplaybreaks
\newcommand{\D}{{\rm{d}}}

\makeatletter
\@addtoreset{equation}{section}
\makeatother

\begin{document}

\begin{center}
{\large {\textbf{Scalar one-loop 4-point integral with one massless vertex in
loop regularization}}}

\vspace{10pt}
Jin Zhang$^{*}$

\vspace{10pt}

\it {School of Physics and Electronic Engineering, Yuxi Normal University\\
Yuxi, Yunnan, 653100, People's Republic of China }

\vspace{10pt}

\vspace{30pt}


\end{center}

\abstract{The scalar one-loop 4-point function with one massless
vertex is evaluated analytically by employing the loop regularization method.
According to the method a characteristic scale $\mu_{s}$ is introduced to regularize the
divergent integrals. The infrared divergent parts, which take the form of
$\ln^{2}(\lambda^{2}/\mu^{2}_{s})$ and
$\ln(\lambda^{2}/\mu^{2}_{s})$ as $\mu_{s}\rightarrow 0$ where
$\lambda$ is a constant and expressed in
terms of masses and Mandelstam variables, and the infrared stable parts
are well separated. The result is shown explicitly via
$44$ dilogarithms in the kinematic sector in which our evaluation is
valid.}

\vfill {\footnotesize $*${E-mail: jinzhang@yxnu.edu.cn}}

\thispagestyle{empty}

\pagebreak

\tableofcontents
\addtocontents{toc}{\protect\setcounter{tocdepth}{2}}



\section{introduction}
The precise tests of physics within the framework of the Standard
Model(SM) of particle physics and finding new physics beyond the SM
always need to
evaluate amplitudes of some physical process at quark level to higher
orders of some coupling constant via perturbation theory. Analytic
results of Feynman diagrams play the key role in investigating the
infrared and ultraviolet structure of a theory but also for ensuing
numerical calculation. Ways approaching this goal involve the multi-loop or/and multi-point Feynman
diagrams evaluation. Up to now the particle physics community has
developed powerful methods for higher orders Feynman diagrams
calculation, techniques in state-of-the-art including
integrating by parts\cite{Chetyrkin:1981qh, Baikov:1996rk}, evaluating by
Mellin-Barnes representation\cite{Boos:1990rg}, differential equations
method\cite{Kotikov:1990kg, Kotikov:1991pm, Kotikov:1991hm, Kotikov:1990zs, Remiddi:1997ny, Argeri:2007up, Henn:2014qga} and so on.
Technical details of each approach and other rare methods,
one can refer to Refs.\cite{smirnov2012feynman, Buttar:2006zd,
Bern:2008ef, Binoth:2010ra} and the references therein. In evaluating Feynman diagrams
we should realize is that there is significant difference in expressing the final results
between massless and massive theories. In massless
theories, the Feynman integrals can be expressed in terms of
polylogarithms\cite{lewindilog, Remiddi:1999ew}. However, the evaluation of massive multiple loop
Feynman integrals are more complicated than the massless cases
since the results can not be expressed via polylogarithms,
the elliptic generalization of polylogarithms, the so-called elliptic polylogarithms,
are needed. Examples in this trend can be found in
Refs.\cite{Bloch:2013tra, Adams:2014vja, Adams:2015gva, Broedel:2017kkb, Broedel:2017siw, Broedel:2019hyg, Remiddi:2017har,
Ellipticinqft2019, Bezuglov:2020ywm},
mathematical ground of elliptic polylogarithms and other properties, in particular the analytic
structures, see Refs.\cite{abeilinson1994, alevin1997, Goncharov:1998kja, alevin0703237, fcsbrown1110.6917, Passarino:2016zcd}.

The evaluation of one-loop integrals of Feynman diagrams holds a prominent position
from both theoretical and experimental sides\cite{Denner:1991kt, Kniehl:1993ay, Bardin:1999ak, Denner:2019vbn}.
It has been shown that the general $N$-point($N\geq 5$) scalar one-loop integrals
can be recursively expressed in terms of combinations of
$(N-1)$-point integrals. Hence an arbitrary $N$-point($N\geq 5$) integral
can be reduced to sum of several scalar one-loop four-point integrals.
Since tensor-type integrals can be reduced to scalar
integrals by the Passarino-Veltman\cite{Passarino:1978jh} scheme,
then we can get the desired results from scalar integrals with
including appropriate tensor structures which
are formed by metric tensor $g_{\mu\nu}$ and external momenta.
Consequently, all types one-loop integrals will be evaluated analytically in principle.
In other words, scalar one-loop four-point integrals play the intermediate role that
transits the intractable $N$-point($N\geq 5$) integrals to accessible ones. Therefore
it is helpful to investigate the scalar four-point integrals carefully.

In the pioneering work by 't Hooft and Veltman\cite{tHooft:1978xw}, scalar
one-loop one-, two-, three- and four-point functions are studied
generally, the scalar four-point function with real masses is
expressed in terms of 24 dilogarithms, but for the case of complex
masses it needs 108 dilogarithms. However, there is still long
way to go before the results can be used in practical applications.
Soon later by using the so-called
projective transformation\cite{tHooft:1978xw}, it is found that the scalar one-loop
four-point function can be reduced to 16 dilogarithms in some
kinematical regions\cite{dennernpb1991}, generalization to
tensor\cite{oldenborgh1990} and to pentagon
integrals\cite{bernnpb1993, bernnpb1994} are also carried out. An
important application is made in Ref.\cite{beenakkernpb1990} which
employe box integrals to study some electroweak processes in SM. A
more complete work is given by Ref.\cite{duplancic2001} which
calculates a set of scalar one-loop four-point integrals with
massless internal lines and some massive external lines, the results
obtained is convenient for analytic continuation.
Scalar one-loop three- and four-point integrals for QCD are calculated
in the space-like region in Ref.\cite{Ellis:2007qk} where the ultraviolet, infrared and
collinear divergent integrals are widely investigated.
A thorough work in evaluating scalar four-point functions,
which are valid for complex masses, is presented
in Ref.\cite{denner2011scalar}, in which all the regular and soft- and/or collinear singular
integrals are analyzed by making use of dimensional and mass regularization.

\begin{figure}
\begin{center}
\includegraphics[scale=0.75]{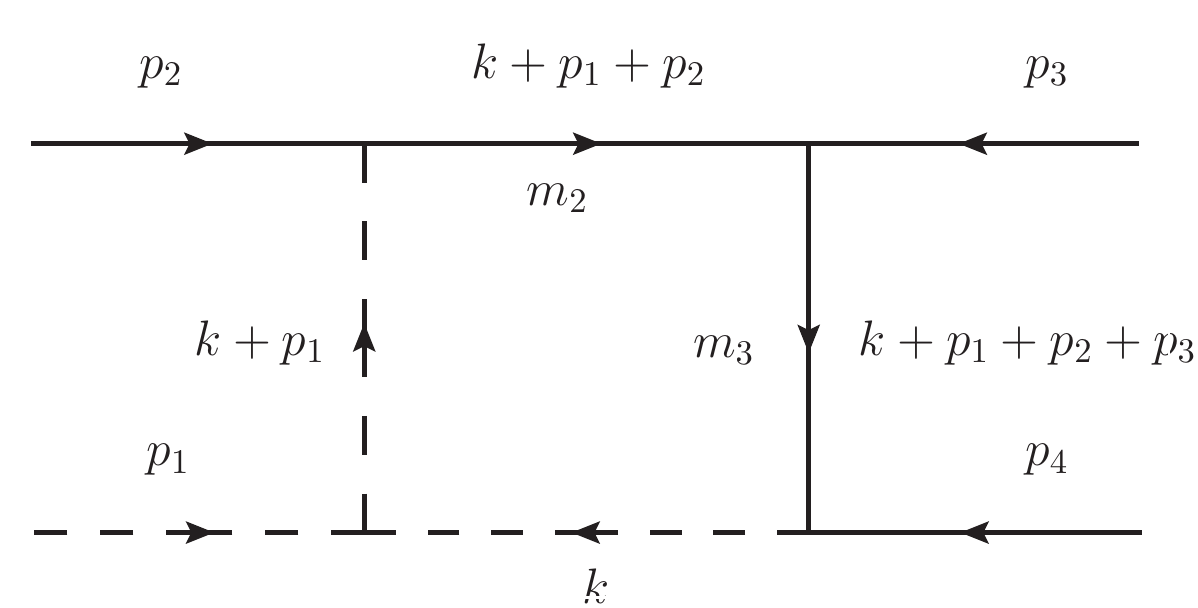}
\caption{Scalar one-loop 4-point integral with one massless vertex.
The dashed and solid lines are massless and massive, respectively.
As usual all the external momenta are
inward.}\label{scalaroneloopbox}
\end{center}
\end{figure}

Nearly all the scalar four-point integrals mentioned above are evaluated
with the dimensional regularization method\cite{tHooft:1972fi}. An alternative
way to extract singular parts in evaluating Feynman
diagrams is loop regularization\cite{Wu:2002xa, Wu:2003dd},
which has been successfully applied in some practical
calculations in hadronic weak decays of $B$ mesons\cite{Su:2010vt, Huang:2011yf, Su:2011eq}.
Motivated by the significant role played by the scalar one-loop
four-point integral in reducing the tedious $N$-point($N\geq 5$)
integrals to tractable ones, in this paper we will evaluate
a typical infrared divergent scalar one-loop four-point integral
as depicted in fig.\ref{scalaroneloopbox} using loop
regularization method. The integral of fig.\ref{scalaroneloopbox},
which corresponds to ``Box $13$" of Ref.\cite{Ellis:2007qk},
is collinear divergent\footnote{A analysis of soft and collinear divergence of
a diagram with loop regularization via Cayley Matrix will be presented in a
separate publication.}. As we know that the divergent structure of a
amplitude is independent of the regularization scheme, but
the expressions of the divergent part and the stable part may be
distinct for different regularization scheme.
Hence the purpose of the paper is, by a specific example, showing
how the infrared divergent and infrared stable parts are extracted via
loop regularization. We stress that in this paper we do not so ambitious
as the aforementioned works on scalar one-loop four-point integrals which try to
investigate the issue under various circumstances thoroughly, we only content ourselves on the
diagram depicted in fig.\ref{scalaroneloopbox}. In this sense
we just present a case study on scalar one loop four-point integral
by loop regularization. We hope that the results shed some
light on the evaluation of scalar one loop four-point
integrals, but also helpful in calculating some box diagram mediated decaying processes.

Before starting our evaluation, the following comments are in order.

(\romannumeral 1)To perform the integrals over
Feynman parameters, the \emph{Euler shift} is adopted. Accordingly,
two equations, i.e., Eq.(\ref{constrainofintegrationone}) and
Eq.(\ref{constrainofbeta}), should be satisfied by the transforming parameters
$\alpha$ and $\beta$. We assume that the two equations have two real roots,
and one root of each equation lies in the range $(0,\,1)$ as stated in
our evaluation. These requirements fix a kinematic allowed sector in the space
spanned by the masses and external momentum, we call it sector \uppercase\expandafter{\romannumeral 1}.

(\romannumeral 2)In our evaluation we need factorize 12 quadratic polynomials of $F$-type which are denoted
by $F_{ij}(i=0,1;\,j=1,2,3)$ and of $G$-type which are denoted by $G_{ij}(i=0,1;\,j=1,2,3)$
into products of their roots. The $F$-type functions are in the denominator
and the $G$-type functions are arguments of logarithms, the coefficients of the two type functions
are formed by on-shell masses and masses of propagators as well as
invariant combination of external momentum.
It is obvious that the factorization should be careful since it depends on if the
quadratic polynomials have two real roots. In other words,
the validity of each factorization determines a kinematic
sector where the quadratic polynomial has two real roots.
Hence in all there are 12 kinematic sectors to be fixed,
and the intersection of them is the
kinematic sector where our evaluation is allowed, we call
it as sector \uppercase\expandafter{\romannumeral 2}. Figuring out
sector \uppercase\expandafter{\romannumeral 2} exactly is difficult
in the complicated space
established by the masses and external momentum.
In order to get rid of the dilemma, we assume that there may be some kinematic sector
in which all the factorizations are valid. A case-by-case
analysis of the similar integrals one can refer to the
appendix of Ref.\cite{Haber:1983fc} and appendix D
of Ref.\cite{npb.323.267}.

The overlap of sector \uppercase\expandafter{\romannumeral 1}
and \uppercase\expandafter{\romannumeral 2} is the desired kinematic
part where the results obtained in this paper can be correctly applied.
We assume that there is some method by which the
overlapping sector can be determined, although we do not find it explicitly
in this paper. It is worth emphasizing that the required kinematic sector
may be unphysical or even not existed, if it were this case, we just present a formal study
on the infrared scalar one-loop four point integral.

The paper is organized as follows. After this short introduction we
display some mathematical functions which are frequently used in our
evaluation in section \uppercase\expandafter{\romannumeral 2}.
Then in Section \uppercase\expandafter{\romannumeral 3}
details of the evaluation and results
are presented. Section \uppercase\expandafter{\romannumeral 4}
contains our short summary. Some necessary
formulae are listed in the appendix.

\section{preliminaries}

We define the following massive scalar one-loop 4-point integral
with one massless vertex, as depicted in Fig.\ref{scalaroneloopbox}
\begin{equation}
I=\int\frac{\D^{4}k}{(2\pi)^{4}}
\frac{1}{\mathcal {D}_{1}\mathcal {D}_{2}\mathcal {D}_{3}\mathcal {D}_{4}},\label{scalarbox}
\end{equation}
where
\begin{eqnarray}
\mathcal {D}_{1}&=&k^{2}+i\epsilon\nonumber\\
\mathcal {D}_{2}&=&(k+p_{1})^{2}+i\epsilon\nonumber\\
\mathcal {D}_{3}&=&(k+p_{1}+p_{2})^{2}
-m_{2}^{2}+i\epsilon\nonumber\\
\mathcal {D}_{4}&=&(k+p_{1}+p_{2}+p_{3})^{2}-m_{3}^{2}+i\epsilon,   \label{fourdenominators}
\end{eqnarray}
The $i\epsilon$ will be systematically retained in our evaluation,
$m_{2}$ and $m_{3}$ are the masses of the two massive internal
lines. As usual we fix that the all
the external momenta are inward, they are related by the energy conservation
\begin{equation}
p_{1}+p_{2}+p_{3}+p_{4}=0,\label{energyconser}
\end{equation}
We assume that the four external momenta satisfy 
\begin{equation}
p^{2}_{1}=0,\quad\quad
p_{2}^{2}=\omega_{2}^{2},\quad\quad p_{3}^{2}=\omega_{3}^{2},\quad
p_{4}^{2}=(p_{1}+p_{2}+p_{3})^{2}=\omega_{4}^{2},  \label{fourexternalmomenta}
\end{equation}
for brevity we define
\begin{equation}
s_{ij}=p_{i}+p_{j}.    \label{combinationsofexternalmomentum}
\end{equation}

The results will be expressed in terms of logarithms and dilogarithms.
As usual we choose the principal value of the logarithms lies in
the negative axis, hence we find
\begin{equation}
\ln(x\pm i\epsilon)=\ln|x|\pm i\pi, \quad\quad x<0  \label{analyticcontoflog}
\end{equation}
In expanding logarithm of products one should take into account
the convention\cite{tHooft:1978xw}
\begin{equation}
\ln(ab)=\ln a+\ln b+\eta(a,b),\label{logprodexpfirst}
\end{equation}
where the $\eta$ term is
\begin{equation}
\eta(a,b)=2\pi
i\{\theta(-{\rm{Im}}\,a)\theta(-{\rm{Im}}\,b)\theta\big({\rm{Im}}\,(ab)\big)
-\theta({\rm{Im}}\,a)\theta({\rm{Im}}\,b)\theta\big(-{\rm{Im}}\,(ab)\big)\}.\label{definingeta}
\end{equation}
following this rule it is easy to get
\begin{gather}
\ln(ab)=\ln a+\ln b,\quad \text{if ${\rm{Im}}(a)$
and ${\rm{Im}}(b)$ have different signs},\nonumber\\
\ln\frac{a}{b}=\ln a-\ln b, \quad \text{if ${\rm{Im}}(a)$ and
${\rm{Im}}(b)$ have the same signs}.\label{invlvoedeta}
\end{gather}

As we know the dilogarithm develops an imaginary part for $x\geq 1$,
then we have\cite{duplancic2001}
\begin{equation}
{\rm{Li}}_{2}(x\pm i \epsilon)=-{\rm{Li}}_{2}\Big(\frac{1}{x}\Big)
-\frac{1}{2}\ln^{2}x+\frac{\pi^{2}}{3}\pm i\pi\ln x,   \label{dilogwithinfiniteimaginery}
\end{equation}
Most of our results involve analytic continuation of dilogarithms of two
variables
\begin{equation}
{\rm{Li}}_{2}(1-x_{1}x_{2}),   \label{dilogoftwovariables}
\end{equation}
this can be completed by the substitution\cite{beenakkernpb1990, denner2011scalar}
\begin{eqnarray}
{\rm{Li}}_{2}(1-x_{1}x_{2})&\rightarrow&
\mathcal {L}(x_{1},x_{2})\nonumber\\
&=&{\rm{Li}}_{2}(1-x_{1}x_{2})+\eta(x_{1},x_{2})\ln(1-x_{1}x_{2}).  \label{dilogwithtwoarguments}
\end{eqnarray}
where $\eta(x_{1},x_{2})$ is given by Eq.(\ref{definingeta}).

\section{calculation and results}
\subsection{basic formula}
Firstly we notice that the first two denominators can be combined
via Feynman parameterization
\begin{equation}
\frac{1}{(k^{2}+i\epsilon)[(k+p_{1})^{2}+i\epsilon]}
=\int^{1}_{0}\D u\,
\frac{1}{[(k+up_{1})^{2}+i\epsilon]^{2}},\,\quad\text{if}\quad
p^{2}_{1}=0   \label{twofeynpara}
\end{equation}
Then primitive integral Eq.(\ref{scalarbox}) becomes
\begin{eqnarray}
I&=&\int\frac{\D^{4}k}{(2\pi)^{4}}
\int^{1}_{0}\D u\,\frac{1}{[(k+up_{1})^{2}+i\epsilon]^{2}}\nonumber\\
&\times&\frac{1}{[(k+p_{1}+p_{2})^{2}-m_{2}^{2}+i\epsilon]
[(k+p_{1}+p_{2}+p_{3})^{2}-m_{3}^{2}+i\epsilon]},\label{rewirtebox}
\end{eqnarray}
Then by using Feynman parameterization twice, Eq.(\ref{rewirtebox}) can
be written in the form
\begin{eqnarray}
I&=&3\,!\int\frac{\D^{4}k}{(2\pi)^{4}}\int^{1}_{0}\D u
\int^{1}_{0}\D x \D y\D z\,\delta(1-x-y-z)\,x\nonumber\\
&\times&\Big\{x\big[(k+up_{1})^{2}+i\epsilon\big]
+y\big[(k+p_{1}+p_{2})^{2}-m_{2}^{2}+i\epsilon\big]\nonumber\\
&+&z\big[(k+p_{1}+p_{2}+p_{3})^{2}-m_{3}^{2}
+i\epsilon\big]\Big\}^{-4},\label{totalfeynpara}
\end{eqnarray}
Sine there is infrared divergence in $I$, in order to
carry out the integration over $k$, an appropriate regularization scheme
must be employed. Instead of the most popular dimensional regularization,
an alternative is loop regularization\cite{Wu:2002xa, Wu:2003dd}.
According to the method, the loop momentum $k$ transforms as
\begin{gather}
k^{2}\rightarrow [k^{2}]_{l}=k^{2}-M_{l}^{2},\nonumber\\
\int\frac{\D^{4}k}{(2\pi)^{4}}
\rightarrow\int\Big[\frac{\D^{4}k}{(2\pi)^{4}}\Big]_{l}
=\lim_{N, M_{i}^{2}
\rightarrow\infty}\sum_{l=0}^{N}c^{N}_{l}\int\frac{d^{4}k}{(2\pi)^{4}}, \label{loopregul}
\end{gather}
which is constrained by
\begin{eqnarray}
\lim_{N, M_{i}^{2}\rightarrow\infty}\sum_{l=0}^{N}c^{N}_{l}(M_{l}^{2})^{2}=0,\quad
c_{0}^{N}=0 \quad (i=0,1,...,N\,\,\text{and}\,\, n=0,1,..).\label{loopconstrain}
\end{eqnarray}
From Eq.(\ref{loopconstrain}) the coefficients $c_{l}^{N}$ can be worked out
\begin{equation}
c_{l}^{N}=(-1)^{l}\frac{N!}{l!(N-l)!},\nonumber
\end{equation}
and the regulator mass is given by
\begin{equation}
M^{2}_{l}=\mu^{2}_{s}+lM^{2}_{R},\label{regulatormass}
\end{equation}
Then it leads to the desired integration form over $k$
\begin{gather}
k^{2}\rightarrow k^{2}-\mu^{2}_{s}-lM^{2}_{R},\nonumber\\
\int\frac{\D^{4}k}{(2\pi)^{4}}\rightarrow
\lim_{N, M_{R}^{2}\rightarrow\infty}\sum_{l=0}^{N}(-1)^{l}
\frac{N!}{l!(N-l)!}\int\frac{\D^{4}k}{(2\pi)^{4}}.\label{loopmomentum}
\end{gather}
If there were only infrared divergence, when the integration
over loop momentum is completed, terms involving $M_{R}$ will vanish
after taking the limit, thus in this case it amounts to
introduce a characteristic scale $\mu_{s}$ in the amplitudes.
With this in mind, after the integration over $k$ is performed, we obtain
\begin{equation}
I=\frac{i}{(4\pi)^{2}}\int_{0}^{1}\D u\int_{0}^{1}
\D x \D y \D z \,x\delta(1-x-y-z) \frac{1}{[Q(u,x,y,z)]^{2}}, \label{loopmoentumkintegrated}
\end{equation}
where $Q$ is defined as
\begin{eqnarray}
Q(u,x,y,z)&=&s_{12}^{2}y^{2}+\omega_{4}^{2}z^{2}+(s_{12}^{2}
+\omega_{4}^{2}-\omega_{3}^{2})yz\nonumber\\
&+&\big[y(s_{12}^{2}-\omega_{2}^{2})+
z(\omega_{4}^{2}-s_{23}^{2})\big]ux
-y(s_{12}^{2}-m_{2}^{2})\nonumber\\
&-&z(\omega_{4}^{2}-m_{3}^{2})-\mu_{s}^{2}-i\epsilon    \label{polynomialQ}
\end{eqnarray}
The integration over $u$ is trivial thus we do it first
\begin{eqnarray}
I=\frac{-i}{(4\pi)^{2}}\int_{0}^{1}\D x \D y \D z\,\delta(1-x-y-z)\frac{1}{B(y,z)}
\Big[\frac{1}{Q(1,x,y,z)}-\frac{1}{Q(0,x,y,z)}\Big],  \label{variablezintegrated}
\end{eqnarray}
where
\begin{equation}
B(y,z)=y(s_{12}^{2}-\omega_{2}^{2})+
z(\omega_{4}^{2}-s_{23}^{2}).       \label{denominatorfuncB}
\end{equation}
The integration over $z$ can be performed immediately, it leads to
\begin{eqnarray}
I&=&\frac{-i}{(4\pi)^{2}}\int_{0}^{1}\D x \int_{0}^{1-x}\D y \,\frac{1}{B(y,1-x-y)}\nonumber\\
&\times&\Big[\frac{1}{Q(1,x,y,1-x-y)}-\frac{1}{Q(0,x,y,1-x-y)}\Big]  \label{feynmannzintegrated}
\end{eqnarray}
To proceed we make the following transformation on $x$ and $y$
\begin{equation}
x=1-x',\quad\quad y=x'-y'
\end{equation}
this yields a convenient form for Eq.(\ref{feynmannzintegrated})
\begin{equation}
I=\frac{-i}{(4\pi)^{2}}\int_{0}^{1}\D x' \int_{0}^{x'}\D y' \,\frac{1}{C(x',y')}
\Big[\frac{1}{W_{1}(x',y')}-\frac{1}{W_{0}(x',y')}\Big],   \label{variablesubxy}
\end{equation}
where
\begin{eqnarray}
C(x',y')&=&(s_{12}^{2}-\omega_{2}^{2})x'
+(s_{13}^{2}-\omega_{3}^{2})y',  \nonumber\\
W_{1}(x',y')&=&\omega_{2}^{2}x'^{2}+\omega_{3}^{2}y'^{2}
+(s_{12}^{2}-\omega_{2}^{2}-\omega_{3}^{2})x'y'
-\omega_{2}^{2}x', \nonumber\\
&+&(m_{3}^{2}-m_{2}^{2}-s_{23}^{2}+\omega_{2}^{2})y'
-\mu_{s}^{2}-i\epsilon, \nonumber\\
W_{0}(x',y')&=&s_{12}^{2}x'^{2}
+\omega_{3}^{2}y'^{2}
+(\omega_{4}^{2}-s_{12}^{2}-\omega_{3}^{2})x'y'\nonumber\\
&-&(s_{12}^{2}-m_{2}^{2})x'
+(s_{12}^{2}-m_{2}^{2}-\omega_{4}^{2}+m_{3}^{2})y'-\mu_{s}^{2}-i\epsilon.   \label{threecomponetsofdenominator}
\end{eqnarray}
For later convenience we split $I$ into two parts
\begin{equation}
I=I_{1}+I_{0},   \label{twocomponetsoforiginalI}
\end{equation}
where the two components are
\begin{eqnarray}
I_{1}&=&\frac{-i}{(4\pi)^{2}}\int_{0}^{1}\D x' \int_{0}^{x'}\D y' \,
\frac{1}{C(x',y')W_{1}(x',y')}\nonumber\\
I_{0}&=&\frac{i}{(4\pi)^{2}}\int_{0}^{1}\D x' \int_{0}^{x'}\D y' \,
\frac{1}{C(x',y')W_{0}(x',y')}  \label{integralspltted}
\end{eqnarray}
We will evaluate $I_{1}$ and $I_{0}$ in the forthcoming two subsections, respectively.

\subsection{evaluation of $I_{1}$}

Since the denominator $W_{1}$ is quadratic both in $x'$ and $y'$,
in order to cope with terms involving $x^{2}$,
we perform the so-called \emph{Euler shift} on $y'$
\begin{equation}
y'=\rho+\alpha x'    \label{firstoneeulershift}
\end{equation}
such that
\begin{equation}
\int_{0}^{1}\D x' \int_{0}^{x'}\D y'
=\int_{0}^{1-\alpha}\D \rho\int_{\rho/(1-\alpha)}^{1}\D x'
-\int_{0}^{-\alpha}\D \rho\int_{-\rho/\alpha}^{1}\D x'  \label{eulersfitperdformed}
\end{equation}
where the parameter $\alpha$ is chosen to obey the condition
\begin{equation}
\omega_{3}^{2}\alpha^{2}+\alpha(s_{23}^{2}
-\omega_{2}^{2}-\omega_{3}^{2})+\omega_{2}^{2}=0 \label{constrainofintegrationone}
\end{equation}
From Eq.(\ref{constrainofintegrationone}) we find
\begin{eqnarray}
\alpha
&=&\frac{1}{2\omega_{3}^{2}}\Big[-(s_{23}^{2}-\omega_{2}^{2}-\omega_{3}^{2})
\pm \lambda^{1/2}(s_{23}^{2},\omega_{2}^{2},\omega_{3}^{2})\Big],  \label{rootsofalpha}
\end{eqnarray}
where $\lambda(x,y,z)$ is the well-known K$\ddot{a}$llen function
\begin{equation}
\lambda(x,y,z)=(x+y+z)^{2}-2xy-2yz-2xz,  \label{kallenfunction}
\end{equation}
In our evaluation we take
\begin{equation}
\alpha=\frac{1}{2\omega_{3}^{2}}\Big[-(s_{23}^{2}-\omega_{2}^{2}-\omega_{3}^{2})
+ \lambda^{1/2}(s_{23}^{2},\omega_{2}^{2},\omega_{3}^{2})\Big]. \label{alphavaluetaken}
\end{equation}
and assume $0<\alpha<1$. According to Eq.(\ref{constrainofintegrationone}),
all the $x'^{2}$-dependent terms vanish,
now $W_{1}$ is linear in $x'$, we denote it as
\begin{eqnarray}
 Y_{1}(x',\rho)=f_{1}(\rho)x'+f_{2}(\rho)  \label{newformofQone}
\end{eqnarray}
where $f_{1}$ and $f_{2}$ only depend on $\rho$, the explicit expressions are
\footnote{Notice that the lower index of $f$ and $g$ tell the maximum power of $\rho$. }
\begin{eqnarray}
f_{1}(\rho)&=&A_{1}\rho+B_{1},\nonumber\\
f_{2}(\rho)&=&A_{2}\rho^{2}+B_{2}\rho-\mu_{s}^{2}-i\epsilon,  \label{twofunctionofonly}
\end{eqnarray}
All the coefficients in Eq.(\ref{twofunctionofonly}) are constants
which are formed by on-shell masses in Eq.(\ref{onshellmasses})
and propagator masses $m_{i}(i=2,3)$ as well as
combinations of external momentum $s_{ij}$ in Eq.(\ref{combinationsofexternalmomentum})
\begin{eqnarray}
A_{1}&=&s_{23}^{2}+(2\alpha-1)\omega_{3}^{2}-\omega_{2}^{2},\nonumber\\
B_{1}&=&(m_{2}^{2}-\omega_{2}^{2})
+\alpha(m_{3}^{2}-m_{2}^{2}+\omega_{2}^{2}-s_{23}^{2}),\nonumber\\
A_{2}&=&\omega_{3}^{2},\nonumber\\
B_{2}&=&m_{3}^{2}-m_{2}^{2}+\omega_{2}^{2}-s_{23}^{2},\nonumber\\
D_{1}&=&s_{13}^{2}-\omega_{3}^{2}.   \label{fourcoefficientsofIone}
\end{eqnarray}
Under the transformation in Eq.(\ref{firstoneeulershift}),
the function $C(x',y')$ besomes
\begin{equation}
X(x',\rho)=g_{0}(\alpha)x'+g_{1}(\rho)  \label{newformofB}
\end{equation}
where $g_{0}(\alpha)$ is constant and $g_{1}(\rho)$ is linear in $\rho$
\begin{equation}
g_{0}(\alpha)
=s_{12}^{2}-\omega_{2}^{2}+\alpha(s_{13}^{2}-\omega_{3}^{2}),\quad
g_{1}(\rho)=D_{1} \rho    \label{cofficientsoffirstintegral}
\end{equation}
This yields a compact form for $I_{1}$, which reads
\begin{eqnarray}
I_{1}&=&\frac{-i}{(4\pi)^{2}}\Big[\int_{0}^{1-\alpha}\D \rho\int_{\rho/(1-\alpha)}^{1}\D x'
-\int_{0}^{-\alpha}\D \rho\int_{-\rho/\alpha}^{1}\D x'\Big]\frac{1}{(g_{0}x'+g_{1})(f_{1}x'+f_{2})}\nonumber\\
&=&\frac{-i}{(4\pi)^{2}}\Big[\int_{0}^{1-\alpha}\D \rho\int_{\rho/(1-\alpha)}^{1}\D x'
-\int_{0}^{-\alpha}\D \rho\int_{-\rho/\alpha}^{1}\D x'\Big]\frac{1}{g_{0}f_{2}-g_{1}f_{1}}\nonumber\\
&\times&\Big(\frac{g_{0}}{g_{0}x'+g_{1}}-\frac{f_{1}}{f_{1}x'+f_{2}}\Big)\nonumber\\
&=&I_{A}+I_{B}   \label{factorizationcompleted}
\end{eqnarray}
where we have split $I_{1}$ into two parts
\begin{eqnarray}
I_{A}&=&\frac{-i}{(4\pi)^{2}}\int_{0}^{1-\alpha}\D \rho\int_{\rho/(1-\alpha)}^{1}\D x'
\frac{1}{g_{0}f_{2}-g_{1}f_{1}}
\Big(\frac{g_{0}}{g_{0}x'+g_{1}}
-\frac{f_{1}}{f_{1}x'+f_{2}}\Big),  \label{integralionea}\\
I_{B}&=&\frac{i}{(4\pi)^{2}}\int_{0}^{-\alpha}\D \rho\int_{-\rho/\alpha}^{1}\D x'
\frac{1}{g_{0}f_{2}-g_{1}f_{1}}
\Big(\frac{g_{0}}{g_{0}x'+g_{1}}
-\frac{f_{1}}{f_{1}x'+f_{2}}\Big).  \label{integralioneb}
\end{eqnarray}
The integration over $x'$ in Eq.(\ref{integralionea}) and Eq.(\ref{integralioneb}) is elementary,
we can carry out it immediately, the results are
\begin{eqnarray}
I_{A}&=&\frac{-i}{(4\pi)^{2}}\int_{0}^{1-\alpha}\D \rho
\frac{1}{g_{0}f_{2}-g_{1}f_{1}}
\Big[\ln\Big(g_{0}+g_{1}\Big)
-\ln\Big(\frac{\rho}{1-\alpha}\,g_{0}+g_{1}\Big)\nonumber\\
&-&\ln\Big(f_{1}+f_{2}\Big)
+\ln\Big(\frac{\rho}{1-\alpha}\,f_{1}+f_{2}\Big)\Big],   \label{xintegratedione}
\end{eqnarray}
and
\begin{eqnarray}
I_{B}&=&\frac{i}{(4\pi)^{2}}\int_{0}^{-\alpha}\D \rho
\frac{1}{g_{0}f_{2}-g_{1}f_{1}}
\Big[\ln\Big(g_{0}+g_{1}\Big)
-\ln\Big(-\frac{\rho}{\alpha}\,g_{0}+g_{1}\Big)\nonumber\\
&-&\ln\Big(f_{1}+f_{2}\Big)
+\ln\Big(-\frac{\rho}{\alpha}\,f_{1}+f_{2}\Big)\Big].   \label{xintegrateditwo}
\end{eqnarray}

Then we divide the upper limit of the integral of $I_{A}$ into two parts
\begin{eqnarray}
I_{1}&=&\frac{-i}{(4\pi)^{2}}\Big(\int_{0}^{1}\D \rho
+\int_{0}^{-\alpha}\D \rho\Big)
\frac{1}{g_{0}f_{2}-g_{1}f_{1}}
\Big[\ln\Big(g_{0}+g_{1}\Big)
-\ln\Big(\frac{\rho}{1-\alpha}\,g_{0}+g_{1}\Big)\nonumber\\
&-&\ln\Big(f_{1}+f_{2}\Big)
+\ln\Big(\frac{\rho}{1-\alpha}\,f_{1}+f_{2}\Big)\Big],   \label{xintegratedionemodified}
\end{eqnarray}
Combining with Eq.(\ref{xintegrateditwo}), after some cancelation we find
\begin{equation}
    I_{1}=\frac{-i}{(4\pi)^{2}}(I_{11}+I_{12}+I_{13})   \label{threecomponetsofIoneone}
\end{equation}
The first component $I_{11}$ is given by
\begin{eqnarray}
I_{11}&=&\int_{0}^{1}\D \rho\,\frac{1}{g_{0}f_{2}-g_{1}f_{1}}
\Big[\ln(g_{0}+g_{1})-\ln(f_{1}+f_{2})\Big]\nonumber\\
&=&\int_{0}^{1}\D \rho
\frac{1}{F_{11}(\rho)}\Big[\ln(g_{0}+g_{1})
-\ln G_{11}(\rho)\Big],        \label{obviousexprofIoneone}
\end{eqnarray}
where
\begin{eqnarray}
F_{11}(\rho)&=&(A_{2}g_{0}-A_{1}D_{1})\rho^{2}
+(B_{2}g_{0}-B_{1}D_{1})\rho-g_{0}(\mu_{s}^{2}+i\epsilon)\nonumber\\
G_{11}(\rho)&=&A_{2}\rho^{2}+(A_{1}+B_{2})\rho+B_{1}-\mu_{s}^{2}-i\epsilon.  \label{fanggofIoneone}
\end{eqnarray}
The other two components are
\begin{eqnarray}
I_{12}&=&\int_{0}^{1-\alpha}\D \rho\,\frac{1}{g_{0}f_{2}-g_{1}f_{1}}
\Big[-\ln\Big(\frac{\rho}{1-\alpha}g_{0}+g_{1}\Big)
+\ln\Big(\frac{\rho}{1-\alpha}f_{1}+f_{2}\Big)\Big],               \label{primaryresultofAtwo} \\
I_{13}&=&\int_{0}^{-\alpha}\D \rho \,\frac{1}{g_{0}f_{2}-g_{1}f_{1}}
\Big[\ln\Big(-\frac{\rho}{\alpha}g_{0}+g_{1}\Big)
-\ln\Big(-\frac{\rho}{\alpha}f_{1}+f_{2}\Big)\Big].  \label{primaryresultofAthree}
\end{eqnarray}
In order to regularize the upper limit of the integral in Eq.(\ref{primaryresultofAtwo},
we make the following variable substitution
\begin{equation}
\rho=(1-\alpha)\xi,   \label{Ionetwosubstituted}
\end{equation}
without confusion, we relabel $\xi$ as $\rho$, then the integral takes the form
\begin{equation}
I_{12}=(1-\alpha)\int_{0}^{1}\D \rho\,\frac{1}{F_{12}(\rho)}
\Big\{-\ln\Big[\Big(g_{0}+(1-\alpha)D_{1}\Big)\rho\Big]
+\ln G_{12}(\rho)\Big\},   \label{newformofIonetwo}
\end{equation}
where
\begin{eqnarray}
F_{12}(\rho)&=&(1-\alpha)^{2}(A_{2}g_{0}-A_{1}D_{1})\rho^{2}
+(1-\alpha)(B_{2}g_{0}-B_{1}D_{1})\rho
-g_{0}(\mu_{s}^{2}+i\epsilon),\nonumber\\
G_{12}(\rho)&=&(1-\alpha)[A_{1}+(1-\alpha)A_{2}]\rho^{2}
+[B_{1}+(1-\alpha)B_{2}]\rho-\mu_{s}^{2}-i\epsilon,  \label{functionsofinteraltwo}
\end{eqnarray}
Similarly, we make the following transformation in Eq.(\ref{primaryresultofAthree})
\begin{equation}
\rho=-\alpha \xi,  \label{Ionethreesubstituted}   
\end{equation}
and relabel $\xi$ as $\rho$, this leads to
\begin{equation}
I_{13}=-\alpha\int_{0}^{1}\D \rho\,
\frac{1}{F_{13}(\rho)}\Big\{\ln\Big[(g_{0}-D_{1} \alpha)\rho\Big]
-\ln G_{13}(\rho)\Big\},  \label{newformofIonetwo}
\end{equation}
where
\begin{eqnarray}
F_{13}(\rho)&=&\alpha^{2}(A_{2}g_{0}-A_{1}D_{1})\rho^{2}
+\alpha(B_{1}D_{1}-B_{2}g_{0})\rho
-g_{0}(\mu_{s}^{2}+i\epsilon)\nonumber\\
G_{13}(\rho)&=&\alpha  (\alpha A_{2} -A_{1})\rho^{2}
-(\alpha B_{2}-B_{1} )\rho-
\mu_{s}^{2}-i\epsilon    \label{functionsofinteralthree}
\end{eqnarray}
The details of evaluating of $I_{11}$, $I_{12}$ and $I_{13}$ are presented in appendix C.

\subsection{ evaluation of $I_{0}$}

The original expression of $I_{0}$ is
\begin{eqnarray}
I_{0}&=&\frac{i}{(4\pi)^{2}}\int_{0}^{1}\D x'
\int_{0}^{x'}\D y' \,\frac{1}{C(x',y')W_{0}(x',y')},   \label{intIzerooriginalform}
\end{eqnarray}
To eliminate the awkward term depending on $x'^{2}$,
we also make the \emph{Euler shift} on $y'$
\begin{equation}
y'=\rho+\beta x'   \label{secondeulersfift}
\end{equation}
such that
\begin{equation}
\int_{0}^{1}\D x' \int_{0}^{x'}\D y'
=\int_{0}^{1-\beta}\D \rho\int_{\rho/(1-\beta)}^{1}\D x'
-\int_{0}^{-\beta}\D \rho\int_{-\rho/\beta}^{1}\D x'  \label{integrationdecompositiion}
\end{equation}
where $\beta$ is chosen to obey the condition
\begin{equation}
\omega_{3}^{2}\beta^{2}+(\omega_{4}^{2}-s_{12}^{2}
-\omega_{3}^{2})\beta+s_{12}^{2}=0   \label{constrainofbeta}
\end{equation}
which renders that all the $x'^{2}$-dependent terms vanish.
The roots of Eq.(\ref{constrainofbeta}) are
\begin{equation}
\beta=\frac{1}{2\omega_{3}^{2}}
\Big[-(\omega_{4}^{2}-s_{12}^{2}-\omega_{3}^{2})
\pm\lambda^{1/2}(\omega_{4}^{2}, s_{12}^{2}, \omega_{3}^{2} )\,\Big]   \label{tworootsofbeta}
\end{equation}
where $\lambda(x, y, z)$ is the Kallen function defined in Eq.(\ref{kallenfunction}).
In our evaluation we take
\begin{equation}
\beta=\frac{1}{2\omega_{3}^{2}}
\Big[-(\omega_{4}^{2}-s_{12}^{2}-\omega_{3}^{2})
+\lambda^{1/2}(\omega_{4}^{2}, s_{12}^{2}, \omega_{3}^{2} )\,\Big] . \label{betavaluetaken}
\end{equation}
and assume $0<\beta<1$. Accordingly, $W_{0}$ is linear in $x'$, we denote it as
\begin{eqnarray}
Y_{0}(x',\rho)=h_{1}(\rho)x'+h_{2}(\rho),   \label{functionyzero}
\end{eqnarray}
where the two functions $h_{1}(\rho)$ and $h_{2}(\rho)$ only depend on $\rho$, they are given by
\begin{eqnarray}
h_{1}(\rho)&=&K_{1}\rho+N_{1}\nonumber\\
h_{2}(\rho)&=&K_{2}\rho^{2}+N_{2}\rho-\mu_{s}^{2}-i\epsilon,    \label{functionhoneandhtwo}
\end{eqnarray}
all the coefficients in Eq.(\ref{functionhoneandhtwo}) are constants
\begin{eqnarray}
K_{1}&=&2(\beta-1)\omega_{3}^{2}+s_{13}^{2}+s_{23}^{2}-\omega_{2}^{2},\nonumber\\
N_{1}&=&(\beta-1)(s_{12}^{2}-m_{2}^{2})-\beta(\omega_{4}^{2}-m_{3}^{2}),\nonumber\\
K_{2}&=&\omega_{3}^{2},\nonumber\\
N_{2}&=&s_{12}^{2}-m_{2}^{2}-\omega_{4}^{2}+m_{3}^{2}.   \label{commoncoefficientsofIzero}
\end{eqnarray}
While $C(x',y')$ transform into
\begin{equation}
X(x',\rho)=g_{0}(\beta)x'+g_{1} (\rho), \label{newformofB}
\end{equation}
where $g_{0}(\beta)$ and $g_{1}(\rho)$ is given by
\begin{eqnarray}
g_{0}(\beta)&=&s_{12}^{2}-\omega_{2}^{2}+\beta(s_{13}^{2}-\omega_{3}^{2}),\nonumber\\
g_{1}(\rho)&=&D_{1} \rho,    \label{cofficientsoffirstintegral}
\end{eqnarray}
Then we rewrite $I_{0}$ in a compact form
\begin{eqnarray}
I_{0}&=&\frac{i}{(4\pi)^{2}}\Big[\int_{0}^{1-\beta}\D \rho\int_{\rho/(1-\beta)}^{1}\D x'
-\int_{0}^{-\beta}\D \rho\int_{-\rho/\beta}^{1}\D x'\Big]
\frac{1}{(g_{0}x'+g_{1})(h_{1}x'+h_{2})}\nonumber\\
&=&\frac{-i}{(4\pi)^{2}}\Big[\int_{0}^{1-\beta}\D \rho\int_{\rho/(1-\beta)}^{1}\D x'
-\int_{0}^{-\beta}\D \rho\int_{-\rho/\beta}^{1}\D x'\Big]\nonumber\\
&\times&\frac{1}{g_{0} h_{2}-g_{1} h_{1}}
\Big(\frac{g_{0}}{ g_{0}x'+g_{1} }-\frac{h_{1}}{ h_{1}x'+h_{2} }\Big).  \label{Izerosplitting}
\end{eqnarray}
After the trivial integration over $x'$ is performed, we find
\begin{equation}
I_{0}=I_{C}+I_{D},   \label{twocomponetsofIzero}
\end{equation}
where
\begin{eqnarray}
I_{C}&=&\frac{i}{(4\pi)^{2}}\int_{0}^{1-\beta}\D \rho\,
\frac{1}{g_{0} h_{2}-g_{1} h_{1}}
\Big[\ln\Big(g_{0}+g_{1}\Big)
-\ln\Big(\frac{\rho}{1-\beta}g_{0}+g_{1}\Big)\nonumber\\
&-&\ln\Big(h_{1}+h_{2}\Big)
+\ln\Big(\frac{\rho}{1-\beta}h_{1}+h_{2}\Big)\Big]\nonumber\\
&=&\frac{-i}{(4\pi)^{2}}\Big(\int_{0}^{1}\D \rho\,+\int_{0}^{-\beta}\D \rho\Big)
\frac{1}{g_{0} h_{2}-g_{1} h_{1}}
\Big[\ln\Big(g_{0}+g_{1}\Big)
-\ln\Big(\frac{\rho}{1-\beta}g_{0}+g_{1}\Big)\nonumber\\
&-&\ln\Big(h_{1}+h_{2}\Big)
+\ln\Big(\frac{\rho}{1-\beta}h_{1}+h_{2}\Big)\Big],  \label{componentICofintIzero}
\end{eqnarray}
and
\begin{eqnarray}
I_{D}&=&\frac{i}{(4\pi)^{2}}\int_{0}^{-\beta}\D \rho\,
\frac{1}{g_{0} h_{2}-g_{1} h_{1}}\nonumber\\
&\times&\Big[\ln(g_{0}+g_{1})
-\ln(-\frac{\rho}{\beta}g_{0}+g_{1})
-\ln(h_{1}+h_{2})+\ln(-\frac{\rho}{\beta}h_{1}+h_{2})\Big].   \label{componentIDofintIzero}
\end{eqnarray}
After some cancelation, we have
\begin{equation}
I_{0}=\frac{i}{(4\pi)^{2}}
(I_{01}+I_{02}+I_{03}), \label{threepiecesofintB}
\end{equation}
The first component is
\begin{eqnarray}
I_{01}&=&\int_{0}^{1}\D \rho\,\frac{1}{g_{0}h_{2}-g_{1}h_{1}}
\Big[\ln(g_{0}+g_{1})-\ln(h_{1}+h_{2})\Big]\nonumber\\
&=&\int_{0}^{1}\D u\,\frac{1}{F_{01}(\rho)}
\Big[\ln\Big(D_{1}u+g_{0}\Big)-\ln G_{01}(\rho)\Big],  \label{firstcomponentIzeroone}
\end{eqnarray}
where
\begin{eqnarray}
F_{01}(\rho)&=&(K_{2}g_{0}-K_{1}D_{1})\rho^{2}
+(g_{0}N_{2}-N_{1}D_{1})\rho-g_{0}(\mu_{s}^{2}+i\epsilon)\nonumber\\
G_{01}(\rho)&=&K_{2}\rho^{2}+(K_{1}+N_{2})\rho+N_{1}-\mu_{s}^{2}-i\epsilon.  \label{FzerooneandGzeroone}
\end{eqnarray}
The second component is
\begin{equation}
I_{02}=\int_{0}^{1-\beta}\D \rho\,\frac{1}{g_{0}h_{2}-g_{1}h_{1}}
\Big[-\ln\Big(\frac{\rho}{1-\beta}g_{0}+g_{1}\Big)
+\ln\Big(\frac{\rho}{1-\beta}h_{1}+h_{2}\Big)\Big],   \label{Izerotwountransfomred}
\end{equation}
We make the following transform on $I_{02}$
\begin{equation}
\rho=(1-\beta)\xi,   \label{Izerotwobyxi}
\end{equation}
and relabel $\xi$ as $\rho$, we obtain
\begin{equation}
I_{02}=(1-\beta)\int_{0}^{1}\D
\rho\,\frac{1}{F_{02}(\rho)}
\Big\{-\ln\Big[\Big(g_{0}+(1-\beta)D_{1}\Big)\rho\Big]
+\ln G_{02}(\rho)\Big\},   \label{Izerotwoxibyrho}
\end{equation}
where
\begin{eqnarray}
F_{02}(\rho)&=&(1-\beta)^{2}(K_{2}g_{0}-K_{1}D_{1})\rho^{2}
-(1-\beta)(N_{1}D_{1}-N_{2}g_{0})\rho-
g_{0}(\mu_{s}^{2}+i\epsilon),\nonumber\\
G_{02}(\rho)&=&(1-\beta )[K_{1}+(1- \beta )K_{2}]\rho^{2}
+[N_{1}+(1- \beta)N_{2}]\rho-\mu_{s}^{2}-i\epsilon.    \label{FzerotwoandGzerotwo}
\end{eqnarray}
The last component is
\begin{equation}
I_{03}=\int_{0}^{-\beta}\D \rho \,\frac{1}{g_{0}h_{2}-g_{1}h_{1}}
\Big[\ln\Big(-\frac{\rho}{\beta}g_{0}+g_{1}\Big)
-\ln\Big(-\frac{u}{\beta}h_{1}+h_{2}\Big)\Big],   \label{componentsintegralofIzero}
\end{equation}
We make the following transformation
\begin{equation}
\rho=-\beta\xi,      \label{Izerothreexi}
\end{equation}
and relabel $\xi$ as $\rho$, the
\begin{equation}
I_{03}=-\beta\int_{0}^{1}\D \rho\,
\frac{1}{F_{03}(\rho)}\Big\{\ln\Big[\Big(g_{0}-\beta D_{1} \Big)\rho\Big]
-\ln G_{03}(\rho)\Big\}     \label{Izerothreexibyrho}
\end{equation}
where
\begin{eqnarray}
F_{03}(\rho)&=&\beta^{2}(K_{2}g_{0}-K_{1}D_{1})\rho^{2}
+\beta(N_{1}D_{1}-N_{2}g_{0})\rho-g_{0}(\mu_{s}^{2}+i\epsilon),\nonumber\\
G_{03}(\rho)&=&\beta(\beta K_{2}-K_{1})\rho^{2}
-(\beta N_{2} -N_{1})\rho-\mu_{s}^{2}-i\epsilon.   \label{FzerothreeandGzerothree}
\end{eqnarray}
The details of evaluating of $I_{01}$, $I_{02}$ and $I_{03}$ are given in appendix D.

\subsection{results and discussions}
Collecting the components in the appendix C and D,
we arrive at the final results for $I$
\begin{equation}
I=I_{1}+I_{2},  \label{twofinalcomponetsofI}
\end{equation}
where the manifest expressions of $I_{1}$ and $I_{1}$ are
\begin{eqnarray}
I_{1}&=&\frac{-i}{(4\pi)^{2}}(I_{11}+I_{12}+I_{13})\nonumber\\
&=&\frac{-i}{(4\pi)^{2}}\frac{1}{a_{1}[\rho^{(1)}_{+}-\rho^{(1)}_{-}]}
\Big\{\ln\frac{\lambda_{1}^{2}}{\mu_{s}^{2}}
+\ln\frac{\lambda_{1}^{2}}{\mu_{s}^{2}}\Big[\ln\alpha-\ln(1-\alpha)\Big]
+\ln\Big|1-\frac{\lambda^{2}}{\mu_{s}^{2}}\Big|
\ln\frac{g_{0}+D_{1}\rho^{(1)}_{+}}{G_{11}(1)}\nonumber\\
&-&\ln\Big|1-\frac{(1-\alpha)\lambda^{2}}{\mu_{s}^{2}}\Big|
\ln\frac{g_{0}+(1-\alpha)D_{1}}{G_{12}(1)}
+\ln\Big|1+\frac{\alpha\lambda_{1}^{2}}{\mu_{s}^{2}}\Big|
\ln\frac{g_{0}-\alpha D_{1}}{G_{13}(1)}
+i\pi\ln\frac{\lambda_{1}^{2}}{\mu_{s}^{2}}\nonumber\\
&+&i\pi\Big[\ln\frac{g_{0}+D_{1}\rho^{(1)}_{+}}{G_{11}(1)}
-\ln\frac{g_{0}+(1-\alpha)D_{1}}{G_{12}(1)}
-\ln(1-\alpha)\Big]
-\frac{\pi^{2}}{6}+\frac{1}{2}\Big[\ln^{2}\alpha-\ln^{2}(1-\alpha)\Big]\nonumber\\
&-&\ln\Big[1-\frac{1}{\rho^{(1)}_{-}}\Big]\ln\frac{g_{0}+D_{1}\rho^{(1)}_{+}}{G_{11}(1)}
+\ln\Big[1-\frac{1-\alpha}{\rho^{(1)}_{-}}\Big]\ln\frac{g_{0}+(1-\alpha)D_{1}}{G_{12}(1)}\nonumber\\
&-&\ln\Big[1+\frac{\alpha}{\rho^{(1)}_{-}}\Big]\ln\frac{g_{0}-\alpha D_{1}}{G_{13}(1)}
+{\rm{Li}}_{2}\Big[\frac{-D_{1}\rho^{(1)}_{+}}{-D_{1}\rho^{(1)}_{+}-g_{0}}\Big]
-{\rm{Li}}_{2}\Big[\frac{D_{1}(1-\rho^{(1)}_{+})}{-D_{1}\rho^{(1)}_{+}-g_{0}}\Big]
\nonumber\\
&-&{\rm{Li}}_{2}\Big[\frac{-D_{1}\rho^{(1)}_{-}}{-D_{1}\rho^{(1)}_{-}-g_{0}}\Big]
+{\rm{Li}}_{2}\Big[\frac{D_{1}(1-\rho^{(1)}_{-})}{-D_{1}\rho^{(1)}_{-}-g_{0}}\Big]\nonumber\\
&+&2{\rm{Li}}_{2}(\frac{\mu_{s}^{2}}{\lambda_{1}^{2}})
+{\rm{Li}}_{2}\Big[1-\frac{\rho^{(1)}_{+}}{1-\rho^{(1)}_{+}}
\frac{\rho^{(1)}_{11}}{1-\rho^{(1)}_{11}}\Big]
+{\rm{Li}}_{2}\Big[1-\frac{\rho^{(1)}_{+}}{1-\rho^{(1)}_{+}}
\frac{\rho^{(2)}_{11}}{1-\rho^{(2)}_{11}}\Big]\nonumber\\
&+&2{\rm{Li}}_{2}\Big[\frac{1}{\rho^{(1)}_{-}}\Big]
-{\rm{Li}}_{2}\Big[1-\frac{\rho^{(1)}_{-}}{1-\rho^{(1)}_{-}}
\frac{\rho^{(1)}_{11}}{1-\rho^{(1)}_{11}}\Big]
-{\rm{Li}}_{2}\Big[1-\frac{\rho^{(1)}_{-}}{1-\rho^{(1)}_{-}}
\frac{\rho^{(2)}_{11}}{1-\rho^{(2)}_{11}}\Big]\nonumber\\
&-&{\rm{Li}}_{2}\Big[\frac{\mu_{s}^{2}}{(1-\alpha)\lambda_{1}^{2}}\Big]
-{\rm{Li}}_{2}\Big[1-\frac{\rho^{(1)}_{+}}{1-\alpha-\rho^{(1)}_{+}}
\frac{\rho^{(1)}_{12}}{1-\rho^{(1)}_{12}}\Big]
-{\rm{Li}}_{2}\Big[1-\frac{\rho^{(1)}_{+}}{1-\alpha-\rho^{(1)}_{+}}
\frac{\rho^{(2)}_{12}}{1-\rho^{(2)}_{12}}\Big]\nonumber\\
&-&{\rm{Li}}_{2}\Big[\frac{1-\alpha}{\rho^{(1)}_{-}}\Big]
+{\rm{Li}}_{2}\Big[1-\frac{\rho^{(1)}_{-}}{1-\alpha-\rho^{(1)}_{-}}
\frac{\rho^{(1)}_{12}}{1-\rho^{(1)}_{12}}\Big]
+{\rm{Li}}_{2}\Big[1-\frac{\rho^{(1)}_{-}}{1-\alpha-\rho^{(1)}_{-}}
\frac{\rho^{(2)}_{12}}{1-\rho^{(2)}_{12}}\Big]\nonumber\\
&+&{\rm{Li}}_{2}\Big[-\frac{\mu_{s}^{2}}{\alpha\lambda_{1}^{2}}\Big]
+{\rm{Li}}_{2}\Big[1+\frac{\rho^{(1)}_{+}}{\alpha+\rho^{(1)}_{+}}
\frac{\rho^{(1)}_{13}}{1-\rho^{(1)}_{13}}\Big]
+{\rm{Li}}_{2}\Big[1+\frac{\rho^{(1)}_{+}}{\alpha+\rho^{(1)}_{+}}
\frac{\rho^{(2)}_{13}}{1-\rho^{(2)}_{13}}\Big]\nonumber\\
&+&{\rm{Li}}_{2}\Big[-\frac{\alpha}{\rho^{(1)}_{-}}\Big]
-{\rm{Li}}_{2}\Big[1+\frac{\rho^{(1)}_{-}}{\alpha+\rho^{(1)}_{-}}
\frac{\rho^{(1)}_{13}}{1-\rho^{(1)}_{13}}\Big]
-{\rm{Li}}_{2}\Big[1+\frac{\rho^{(1)}_{-}}{\alpha+\rho^{(1)}_{-}}
\frac{\rho^{(2)}_{13}}{1-\rho^{(2)}_{13}}\Big]\Big\},  \label{finalresultofcomponentA}
\end{eqnarray}
and
\begin{eqnarray}
I_{0}&=&\frac{i}{(4\pi)^{2}}(I_{01}+I_{02}+I_{03})\nonumber\\
&=&\frac{i}{(4\pi)^{2}}\frac{1}{a_{0}[\rho^{(0)}_{+}-\rho^{(0)}_{-}]}
\Big\{\ln^{2}\frac{\lambda_{0}^{2}}{\mu_{s}^{2}}
+\ln\frac{\lambda_{0}^{2}}{\mu_{s}^{2}}\Big[\ln\beta-\ln(1-\beta)\Big]
+\ln\Big|1-\frac{\lambda_{0}^{2}}{\mu_{s}^{2}}\Big|
\ln\frac{g_{0}+D_{1}\rho_{+}^{(0)}}{G_{01}(1)}\nonumber\\
&-&\ln\Big|1-\frac{(1-\beta)\lambda_{0}^{2}}{\mu_{s}^{2}}\Big|
\ln\frac{g_{0}+(1-\beta)D_{1}}{G_{02}(1)}
+\ln\Big|1+\frac{\beta\lambda_{0}^{2}}{\mu_{s}^{2}}\Big|
\ln\frac{g_{0}-\beta D_{1}}{G_{03}(1)}\nonumber\\
&+&i\pi\ln\frac{\lambda_{0}^{2}}{\mu_{s}^{2}}
+i\pi\Big[\ln\frac{g_{0}+D_{1}\rho_{+}^{(0)}}{G_{01}(1)}
-\ln\frac{g_{0}+(1-\beta)D_{1}}{G_{02}(1)}
-\ln(1-\beta)\Big]\nonumber\\
&-&\frac{\pi^{2}}{6}+\frac{1}{2}\Big[\ln^{2}\beta-\ln^{2}(1-\beta)\Big]
-\ln\Big(1-\frac{1}{\rho_{-}^{(0)}}\Big)
\ln\frac{g_{0}+D_{1}\rho_{-}^{(0)}}{G_{01}(1)}\nonumber\\
&+&\ln\Big(1-\frac{1-\beta}{\rho_{-}^{(0)}}\Big)
\ln\frac{g_{0}+(1-\beta)D_{1}}{G_{02}(1)}
-\ln\Big(1+\frac{\beta}{\rho_{-}^{(0)}}\Big)
\ln\frac{g_{0}-\beta D_{1}}{G_{03}(1)}\nonumber\\
&+&{\rm{Li}}_{2}\Big[\frac{-D_{1}\rho^{(0)}_{+}}{-D_{1}\rho^{(0)}_{+}-g_{0}}\Big]
-{\rm{Li}}_{2}\Big[\frac{D_{1}(1-\rho^{(0)}_{+})}{-D_{1}\rho^{(0)}_{+}-g_{0}}\Big]\nonumber\\
&-&{\rm{Li}}_{2}\Big[\frac{-D_{1}\rho^{(0)}_{-}}{-D_{1}\rho^{(0)}_{-}-g_{0}}\Big]
+{\rm{Li}}_{2}\Big[\frac{D_{1}(1-\rho^{(0)}_{-})}{-D_{1}\rho^{(0)}_{-}-g_{0}}\Big]\nonumber\\
&+&2{\rm{Li}}_{2}(\frac{\mu_{s}^{2}}{\lambda_{0}^{2}})
+{\rm{Li}}_{2}\Big[1-\frac{\rho^{(0)}_{+}}{1-\rho^{(0)}_{+}}
\frac{\rho^{(1)}_{01}}{1-\rho^{(1)}_{01}}\Big]
+{\rm{Li}}_{2}\Big[1-\frac{\rho^{(0)}_{+}}{1-\rho^{(0)}_{+}}
\frac{\rho^{(2)}_{01}}{1-\rho^{(2)}_{01}}\Big]\nonumber\\
&+&2{\rm{Li}}_{2}\Big[\frac{1}{\rho^{(0)}_{-}}\Big]
-{\rm{Li}}_{2}\Big[1-\frac{\rho^{(0)}_{-}}{1-\rho^{(0)}_{-}}
\frac{\rho^{(1)}_{01}}{1-\rho^{(1)}_{01}}\Big]
-{\rm{Li}}_{2}\Big[1-\frac{\rho^{(0)}_{-}}{1-\rho^{(0)}_{-}}
\frac{\rho^{(2)}_{01}}{1-\rho^{(2)}_{01}}\Big]\nonumber\\
&-&{\rm{Li}}_{2}\Big[\frac{\mu_{s}^{2}}{(1-\beta)\lambda_{0}^{2}}\Big]
-{\rm{Li}}_{2}\Big[1-\frac{\rho^{(0)}_{+}}{1-\beta-\rho^{(0)}_{+}}
\frac{\rho^{(1)}_{02}}{1-\rho^{(1)}_{02}}\Big]
-{\rm{Li}}_{2}\Big[1-\frac{\rho^{(0)}_{+}}{1-\beta-\rho^{(0)}_{+}}
\frac{\rho^{(2)}_{02}}{1-\rho^{(2)}_{02}}\Big]\nonumber\\
&-&{\rm{Li}}_{2}\Big[\frac{1-\beta}{\rho^{(0)}_{-}}\Big]
+{\rm{Li}}_{2}\Big[1-\frac{\rho^{(0)}_{-}}{1-\beta-\rho^{(0)}_{-}}
\frac{\rho^{(1)}_{02}}{1-\rho^{(1)}_{02}}\Big]
+{\rm{Li}}_{2}\Big[1-\frac{\rho^{(0)}_{-}}{1-\beta-\rho^{(0)}_{-}}
\frac{\rho^{(2)}_{02}}{1-\rho^{(2)}_{02}}\Big]\nonumber\\
&+&{\rm{Li}}_{2}\Big(-\frac{\mu_{s}^{2}}{\beta\lambda_{0}^{2}}\Big)
+{\rm{Li}}_{2}\Big[1+\frac{\rho^{(0)}_{+}}{\beta+\rho^{(0)}_{+}}
\frac{\rho^{(1)}_{03}}{1-\rho^{(1)}_{03}}\Big]
+{\rm{Li}}_{2}\Big[1+\frac{\rho^{(0)}_{+}}{\beta+\rho^{(0)}_{+}}
\frac{\rho^{(2)}_{03}}{1-\rho^{(2)}_{03}}\Big]\nonumber\\
&+&{\rm{Li}}_{2}\Big[-\frac{\beta}{\rho^{(0)}_{-}}\Big]
-{\rm{Li}}_{2}\Big[1+\frac{\rho^{(0)}_{-}}{\beta+\rho^{(0)}_{-}}
\frac{\rho^{(1)}_{03}}{1-\rho^{(1)}_{03}}\Big]
-{\rm{Li}}_{2}\Big[1+\frac{\rho^{(0)}_{-}}{\beta+\rho^{(0)}_{-}}
\frac{\rho^{(2)}_{03}}{1-\rho^{(2)}_{03}}\Big]\Big\}.  \label{finalresultofcomponentB}
\end{eqnarray}
Combining Eq.(\ref{finalresultofcomponentA}) and Eq.(\ref{finalresultofcomponentB})
we get the final result for the integral in Eq.(\ref{scalarbox}).
In dimensional regularization scheme, the
infrared divergence in Eq.(\ref{scalarbox}) can be expressed in a generic form as
\begin{equation}
\frac{A}{\epsilon^{2}_{{\rm{IR}}}}
+\frac{B}{\epsilon_{{\rm{IR}}}}, \quad
\epsilon_{{\rm{IR}}}=d-4     \label{divergenceofdimreg}
\end{equation}
where $d$ is the dimension of space-time, coefficient $A$ and $B$ are complex functions
of kinematic variables. While in loop regularization scheme,
the infrared divergence appears as terms proportional $\ln^{2}(\lambda_{i}^{2}/\mu_{s}^{2})$
or $\ln(\lambda_{i}^{2}/\mu_{s}^{2})(i=0,\,1)$ when the characteristic scale $\mu_{s}\rightarrow 0$.
There are infrared divergent and finite imaginary terms, they
are generated by analytic continuation of logarithms and dilogarithms,
i.e., Eq.(\ref{analyticcontoflog}) and Eq.(\ref{dilogwithinfiniteimaginery}), respectively.
Since we hold the $i\epsilon$ systematically in the evaluation,
all the dilogarithms with two arguments of the form
\begin{equation}
{\rm{Li}}_{2}(1-x_{1}x_{2}).  \label{dilogoftwoargumentfinal}
\end{equation}
can be expedient to make analytic continuation via Eq.(\ref{dilogwithtwoarguments}).
But for brevity we retain the original expressions for every
term like Eq.(\ref{dilogoftwoargumentfinal}).

\section{conclusions}
In this paper the scalar one-loop 4-point function with a massless
vertex are calculated analytically by loop regularization,
infrared divergent and stable parts are well separated.
The results may be convenient to analytically continue to other
kinematic sectors which is beyond our assumption.
Following the steps adopted in this paper, we may evaluate one loop
tensor-type four-point integrals. We hope the results
obtained in this paper are helpful for evaluating some box mediated processes.

\section{acknowledgements}
The author thanks H. E. Haber for helpful
discussion on the properties of dilogarithms.

\appendix

\section{ Factorization of quadratic equation with two real roots}

Suppose $f(x)$ is quadratic polynomial with imaginary part
\begin{equation}
f(x)=ax^{2}+bx-c(\mu_{s}^{2}+i\epsilon),
\quad\quad  b>0,\quad b^{2}-4ac\mu_{s}^{2}>0    \label{quadraticequ}
\end{equation}
where $\epsilon$ is real positive infinitesimal. The two zeros of Eq.(\ref{quadraticequ}) are
\begin{eqnarray}
x&=&\frac{1}{2a}\Big[-b\pm\sqrt{b^{2}+4ac(\mu_{s}^{2}+i\epsilon)}\,\Big]\nonumber\\
&=&\frac{1}{2a}\Big\{-b\pm\sqrt{b^{2}-4ac\mu_{s}^{2}}
\Big[1+2ac\cdot i\epsilon+\mathcal {O}(\epsilon^{2})+...\Big]\Big\}\nonumber\\
&\approx&\frac{1}{2a}\Big[-b\pm\sqrt{b^{2}-4ac\mu_{s}^{2}}\pm 2ac\cdot i\epsilon\Big],  \label{twozerosofquadpoly}
\end{eqnarray}
where we have use the property that the product of any finite quantity with $\epsilon$ is still a real
infinitesimal, we still denote it by $\epsilon$.
Suppose $\mu_{s}^{2}<<1 $ then we may expand the two zeros as power series of $\mu_{s}^{2}$
\begin{equation}
x=\frac{1}{2a}\Big\{-b\pm b\Big[1-\frac{2ac}{b^{2}}\mu_{s}^{2}+\mathcal {O}(\mu_{s}^{4})\Big]
\pm 2ac\cdot i\epsilon\Big\},  \label{expansionoftwozeros}
\end{equation}
then we get two roots
\begin{eqnarray}
x^{(+)}&=&-\frac{c }{b}\mu_{s}^{2}+i\epsilon, \nonumber\\
x^{(-)}&=&-\frac{b}{a}+\frac{c}{b}\mu_{s}^{2}-i\epsilon,   \label{explicitexproftwozers}
\end{eqnarray}
Now we would like to factorize Eq.(\ref{quadraticequ}) as
\begin{equation}
f(x)=a[x-x^{(+)}][x-x^{(-)}],   \label{quadraticfactorizing}
\end{equation}
then we obtain
\begin{equation}
a=\frac{f(1)}{[1-x^{(+)}][1-x^{(-)}]},  \label{reexpressingofa}
\end{equation}
This leads to an useful factorization on $f(x)$
\begin{equation}
f(x)=\frac{f(1)[x-x^{(+)}][x-x^{(-)}]}
{[1-x^{(+)}][1-x^{(-)}]}, \label{newformoffactorizing}
\end{equation}
assuming $f(x)$ is real and $f(1)>0$, according to Eq.(\ref{invlvoedeta}),
we find useful expression below
\begin{equation}
\ln f(x)=\ln f(1) +\ln\frac{[x-x^{(+)}][x-x^{(-)}]}
{[1-x^{(+)}][1-x^{(-)}]}.  \label{factorizationoflog}
\end{equation}

\section{Useful auxiliary integrals}

In this section we list some integral formula which are useful in the calculation,
they are taken from Refs.\cite{devoto1984, Ellis:2007qk}.
\begin{eqnarray}
&&\int_{0}^{1}\D x\frac{\ln x}{a+bx}
=\frac{1}{b}{\rm{Li}}_{2}\Big(-\frac{b}{a}\Big),   \label{auxiliaryformulaone}
 \\
&&\int_{0}^{1}\D x\frac{\ln(c+ex)}{a+bx}
=\frac{1}{b}\Big\{\ln\Big(\frac{bc-ae}{b}\Big)\ln\frac{a+b}{a}\nonumber\\
&&-{\rm{Li}}_{2}\Big[\frac{e(a+b)}{ae-bc}\Big]
+{\rm{Li}}_{2}\Big(\frac{ae}{ae-bc}\Big)\Big\},      \label{auxiliaryformulatwo}
 \\
&&\int_{0}^{1}\D x\frac{1}{x-x_{0}}\ln\Big[\frac{(x-x_{1})(x-x_{2})}{(1-x_{1})(1-x_{2})}\Big]\nonumber\\
&&=-\,{\rm{Li}}_{2}\Big(1-\frac{x_{0}-1}{x_{0}}\frac{x_{1}}{x_{1}-1}\Big)
-{\rm{Li}}_{2}\Big(1-\frac{x_{0}-1}{x_{0}}\frac{x_{2}}{x_{2}-1}\Big)\nonumber\\
&&+\,{\rm{Li}}_{2}\Big(\frac{1}{1-x_{1}}\Big)
+{\rm{Li}}_{2}\Big(\frac{1}{1-x_{2}}\Big)
+2{\rm{Li}}_{2}\Big(\frac{1}{x_{0}}\Big),  \label{auxiliaryformulathree}
\end{eqnarray}

\section{evaluating the three components of $I_{1}$}
In this section we present the details of evaluation of
the three component of $I_{1}$. In appendix C and appendix D,
Eq.(\ref{dilogwithinfiniteimaginery}) is frequently employed.
In principle, dilogarithms with two arguments can be analytically continued by using
Eq.(\ref{dilogwithtwoarguments}), but for brevity we do not make the analytic
continuation and just retain the original form of them.

\begin{itemize}
\item evaluation of $I_{11}$
\end{itemize}
The integral is
\begin{eqnarray}
I_{11}&=&\int_{0}^{1}\D \rho\,\frac{1}{g_{0}f_{2}-g_{1}f_{1}}
\Big[\ln(g_{0}+g_{1})-\ln(f_{1}+f_{2})\Big]\nonumber\\
&=&\int_{0}^{1}\D \rho\,
\frac{1}{F_{11}(\rho)}
\Big[\ln(D_{1}\rho+g_{0})-\ln G_{11}(\rho)\Big],   \label{intIoneonesimplified}
\end{eqnarray}
where the denominator $F_{11}(\rho)$ is quadratic in $\rho$
\begin{eqnarray}
F_{11}(\rho)&=&
(A_{2}g_{0}-A_{1}D_{1})\rho^{2}
+(B_{2}g_{0}-B_{1}D_{1})\rho-g_{0}(\mu_{s}^{2}+i\epsilon)\nonumber\\
&=&a_{11}[\rho-\rho^{(+)}_{11}][\rho-\rho^{(-)}_{11}],  \label{factorizrionoffoneone}
\end{eqnarray}
The two zeros in Eq.(\ref{factorizrionoffoneone}) are
\begin{eqnarray}
a_{11}=a_{1}&=&A_{2}g_{0}-A_{1}D_{1},\nonumber\\
\rho^{(+)}_{11}=\rho^{(+)}_{1}&=&\frac{\mu_{s}^{2}}{\lambda_{1}^{2}}+i\epsilon,\quad\quad
\rho^{(-)}_{11}=\rho^{(-)}_{1}=-\Big(\kappa_{1}+\frac{\mu_{s}^{2}}{\lambda_{1}^{2}}\Big)-i\epsilon, \label{twozerosofintIoneone}
\end{eqnarray}
where in order to simplifying the symbols, we have
relabel $a_{11}$, $\rho^{(+)}_{11}$ and $\rho^{(-)}_{11}$ as
$a_{1}$, $\rho^{(+)}_{1}$ and $\rho^{(-)}_{1}$, respectively.
The dimensional $\lambda_{1}$ and dimensionless $\kappa_{1}$ are defined as follows
\begin{eqnarray}
\lambda_{1}=\frac{B_{2}g_{0}-B_{1}D_{1}}{g_{0}},\quad\quad
\kappa_{1}=\frac{B_{2}g_{0}-B_{1}D_{1}}{A_{2}g_{0}-A_{1}D_{1}}, \label{lambdaoneandkappaone}
\end{eqnarray}
The manifest expression of $G_{11}$ is
\begin{equation}
G_{11}(\rho)=A_{2}\rho^{2}+(A_{1}+B_{2})\rho+B_{1}-\mu_{s}^{2}-i\epsilon,   \label{functionGoneone}
\end{equation}
By employing the equations in appendix A, it is not difficult to factorize $G_{11}$
into the product of its roots
\begin{equation}
G_{11}(\rho)=\frac{G_{11}(1)\big[\rho-\rho^{(1)}_{11}\big]\big[\rho-\rho^{(2)}_{11}\big]}
{\big[1-\rho^{(1)}_{11}\big]\big[1-\rho^{(2)}_{11}\big]}, \label{functionGoneonefactorized}   
\end{equation}
where
\begin{eqnarray}
G_{11}(1)&=&m_{3}^{2}-\omega_{2}^{2}
+\alpha(m_{3}^{2}-m_{2}^{2}-\omega_{2}^{2}+s_{23}^{2})-\mu_{s}^{2}-i\epsilon\nonumber\\
\rho^{(1)}_{11}&=&\frac{1}{2A_{2}}\big[-(A_{1}+B_{2})+\sqrt{\Delta_{11}}\big]\nonumber\\
\rho^{(2)}_{11}&=&\frac{1}{2A_{2}}\big[-(A_{1}+B_{2})-\sqrt{\Delta_{11}}\big]\nonumber\\
\Delta_{11}&=&(A_{1}+B_{2})^{2}-4A_{2}(B_{1}-\mu_{s}^{2}-i\epsilon),  \label{zerosofGoneone}
\end{eqnarray}
By using the formula in appendix B, we obtain
\begin{eqnarray}
I_{11}&=&\int_{0}^{1}\D \rho\,
\frac{1}{a_{1}[\rho-\rho^{(+)}_{1}][\rho-\rho^{(-)}_{1}]}
\Big\{\ln\Big(D_{1}\rho+g_{0}\Big)
-\ln\frac{G_{11}(1)\big[\rho-\rho^{(1)}_{11}\big]\big[\rho-\rho^{(2)}_{11}\big]}
{\big[1-\rho^{(1)}_{11}\big]\big[1-\rho^{(2)}_{11}\big]}\Big\}\nonumber\\
&=&\frac{1}{a_{1}[\rho^{(+)}_{1}-\rho^{(-)}_{1}]}
\int_{0}^{1}\D \rho\,
\Big[\frac{1}{\rho-\rho^{(+)}_{1}}-\frac{1}{\rho-\rho^{(-)}_{1}}\Big]\nonumber\\
&\times&\Big\{\ln\Big(D_{1}\rho+g_{0}\Big)
-\ln\frac{G_{11}(1)}{\mu^{2}}
-\ln\frac{\big[\rho-\rho^{(1)}_{11}\big]\big[\rho-\rho^{(2)}_{11}\big]}
{\big[1-\rho^{(1)}_{11}\big]\big[1-\rho^{(2)}_{11}\big]}\Big\}\nonumber\\
&=&\frac{1}{a_{1}[\rho^{(+)}_{1}-\rho^{(-)}_{1}]}
\Big\{\ln^{2}\frac{\mu_{s}^{2}}{\lambda_{1}^{2}}
+\ln\Big|1-\frac{\lambda_{1}^{2}}{\mu_{s}^{2}}\Big|
\ln\frac{g_{0}+D_{1}\rho^{(+)}_{1}}{G_{1}(1)}
+2i\pi\ln\frac{\lambda_{1}^{2}}{\mu_{s}^{2}}\nonumber\\
&+&i\pi\ln\frac{g_{0}+D_{1}\rho^{(+)}_{1}}{G_{1}(1)}-\frac{2\pi^{2}}{3}
-\ln\Big[1-\frac{1}{\rho^{(-)}_{1}}\Big]\ln\frac{g_{0}+D_{1}\rho^{(-)}_{1}}{G_{1}(1)}\nonumber\\
&+&{\rm{Li}}_{2}\Big[\frac{-D_{1}\rho^{(+)}_{1}}{-D_{1}\rho^{(+)}_{1}-g_{0}}\Big]
-{\rm{Li}}_{2}\Big[\frac{D_{1}(1-\rho^{(+)}_{1})}{-D_{1}\rho^{(+)}_{1}-g_{0}}\Big]
\nonumber\\
&-&{\rm{Li}}_{2}\Big[\frac{-D_{1}\rho^{(-)}_{1}}{-D_{1}\rho^{(-)}_{1}-g_{0}}\Big]
+{\rm{Li}}_{2}\Big[\frac{D_{1}(1-\rho^{(-)}_{1})}{-D_{1}\rho^{(-)}_{1}-g_{0}}\Big]\nonumber\\
&+&2{\rm{Li}}_{2}\Big(\frac{\mu_{s}^{2}}{\lambda_{1}^{2}}\Big)
+{\rm{Li}}_{2}\Big[1-\frac{\rho^{(+)}_{1}}{1-\rho^{(+)}_{1}}
\frac{\rho^{(1)}_{11}}{1-\rho^{(1)}_{11}}\Big]
+{\rm{Li}}_{2}\Big[1-\frac{\rho^{(+)}_{1}}{1-\rho^{(+)}_{1}}
\frac{\rho^{(2)}_{11}}{1-\rho^{(2)}_{11}}\Big]\nonumber\\
&+&2{\rm{Li}}_{2}\Big[\frac{1}{\rho^{(-)}_{1}}\Big]
-{\rm{Li}}_{2}\Big[1-\frac{\rho^{(-)}_{1}}{1-\rho^{(-)}_{1}}
\frac{\rho^{(1)}_{11}}{1-\rho^{(1)}_{11}}\Big]
-{\rm{Li}}_{2}\Big[1-\frac{\rho^{(-)}_{1}}{1-\rho^{(-)}_{1}}
\frac{\rho^{(2)}_{11}}{1-\rho^{(2)}_{11}}\Big]\Big\}  \label{IoneAfinal}
\end{eqnarray}

\begin{itemize}
\item evaluation of $I_{12}$
\end{itemize}
After the variable substitution, the integral is
\begin{eqnarray}
I_{12}&=&\int_{0}^{1-\alpha}\D \rho\,\frac{1}{g_{0}f_{2}-g_{1}f_{1}}
\Big[-\ln\Big(\frac{\rho}{1-\alpha}g_{0}+g_{1}\Big)
+\ln\Big(\frac{\rho}{1-\alpha}f_{1}+f_{2}\Big)
\Big]\nonumber\\
&=&(1-\alpha)\int_{0}^{1}\D \rho\,\frac{1}{F_{12}(\rho)}
\Big\{-\ln\Big[\Big(g_{0}+(1-\alpha)D_{1}\Big)\rho\Big]
+\ln G_{12}(\rho)\Big\},    \label{intIonetwovariablesub}
\end{eqnarray}
where $F_{12}(\rho)$ is given by
\begin{eqnarray}
F_{12}(\rho)&=&(1-\alpha)^{2}(A_{2}g_{0}-A_{1}D_{1})\rho^{2}
+(1-\alpha)(B_{2}g_{0}-B_{1}D_{1})\rho
-g_{0}(\mu_{s}^{2}+i\epsilon)\nonumber\\
&=&a_{12}[\rho-\rho^{(+)}_{12}][\rho-\rho^{(-)}_{12}], \label{functionFonetwoofintonetwo}
\end{eqnarray}
with
\begin{equation}
a_{12}=(1-\alpha)^{2}(A_{2}g_{0}-A_{1}D_{1}),\quad
\rho^{(+)}_{12}=\frac{\rho^{(+)}_{1}}{1-\alpha}, \quad
\rho^{(-)}_{12}=\frac{\rho^{(-)}_{1}}{1-\alpha},\label{functionFonetwofactorized}
\end{equation}
The argument of the second logarithm in the numerator is defined as
\begin{eqnarray}
G_{12}(\rho)&=&(1-\alpha)[A_{1}+(1-\alpha)A_{2}]\rho^{2}
+[B_{1}+(1-\alpha)B_{2}]\rho-\mu_{s}^{2}-i\epsilon\nonumber\\
&=&\frac{G_{12}(1)\big[\rho-\rho^{(1)}_{12}\big]\big[\rho-\rho^{(2)}_{12}\big]}
{\big[1-\rho^{(1)}_{12}\big]\big[1-\rho^{(2)}_{12}\big]}, \label{functionGontwo}
\end{eqnarray}
with
\begin{eqnarray}
G_{12}(1)&=&m_{3}^{2}-\mu_{s}^{2}-i\epsilon\nonumber\\
\rho^{(1)}_{12}&=&\frac{1}{2(1-\alpha)[A_{1}+(1-\alpha)A_{2}]}
\{-[B_{1}+(1-\alpha)B_{2}]+\sqrt{\Delta_{12}}\}\nonumber\\
\rho^{(2)}_{12}&=&\frac{1}{2(1-\alpha)[A_{1}+(1-\alpha)A_{2}]}
\{-[B_{1}+(1-\alpha)B_{2}]-\sqrt{\Delta_{12}}\}\nonumber\\
\Delta_{12}&=&[B_{1}+(1-\alpha)B_{2}]^{2}
+4(1-\alpha)[A_{1}+(1-\alpha)A_{2}](\mu_{s}^{2}-i\epsilon), \label{functionGonetwofactorized}
\end{eqnarray}
where Eq.(\ref{constrainofintegrationone}) has been employed.
By using the equation in appendix B, it is not difficult to get
\begin{eqnarray}
I_{12}&=&(1-\alpha)\int_{0}^{1}\D \rho\,
\frac{1}{a_{12}[\rho-\rho^{(+)}_{12}][\rho-\rho^{(-)}_{12}]}\nonumber\\
&\times&\Big\{-\ln\Big[\Big(g_{0}+(1-\alpha)D_{1}\Big)\rho\Big]
+\ln\frac{G_{12}(1)[\rho-\rho^{(1)}_{12}][\rho-\rho^{(2)}_{12}]}
{[1-\rho^{(1)}_{12}][1-\rho^{(2)}_{12}]}\,\Big\}\nonumber\\
&=&\frac{(1-\alpha)}{a_{12}[\rho^{(+)}_{12}-\rho^{(-)}_{12}]}
\int_{0}^{1}\D \rho\,
\Big[\frac{1}{\rho -\rho^{(+)}_{12}}
-\frac{1}{\rho-\rho^{(-)}_{12}}\Big]\nonumber\\
&\times&\Big\{-\ln\Big[\Big(g_{0}+(1-\alpha)D_{1}\Big)\rho\Big]
+\ln\frac{G_{12}}{\mu^{2}}+
\ln\frac{[\rho-\rho^{(1)}_{12}][\rho-\rho^{(2)}_{12}]}
{[1-\rho^{(1)}_{12}][1-\rho^{(2)}_{12}]}\,\Big\}\nonumber\\
&=&\frac{1}{a_{1}[\rho^{(+)}_{1}-\rho^{(-)}_{1}]}
\Big\{-\frac{1}{2}\ln^{2}\frac{\lambda_{1}^{2}}{\mu_{s}^{2}}
-\ln(1-\alpha)\ln\frac{\lambda_{1}^{2}}{\mu_{s}^{2}}
-\frac{1}{2}\ln^{2}(1-\alpha)\nonumber\\
&-&\ln\Big|1-\frac{(1-\alpha)\lambda_{1}^{2}}{\mu_{s}^{2}}\Big|
\ln\frac{g_{0}+(1-\alpha)D_{1}}{G_{12}(1)}
-i\pi\ln\frac{\lambda_{1}^{2}}{\mu_{s}^{2}}
-i\pi\ln(1-\alpha)\nonumber\\
&-&i\pi\ln\frac{g_{0}+(1-\alpha)D_{1}}{G_{12}(1)}+\frac{\pi^{2}}{3}
+\ln\Big[1-\frac{1-\alpha}{\rho^{(-)}_{1}}\Big]
\ln\frac{g_{0}+(1-\alpha)D_{1}}{G_{12}(1)}\nonumber\\
&-&{\rm{Li}}_{2}\Big[\frac{\mu_{s}^{2}}{(1-\alpha)\lambda_{1}^{2}}\Big]
-{\rm{Li}}_{2}\Big[1-\frac{\rho^{(+)}_{1}}{1-\alpha-\rho^{(+)}_{1}}
\frac{\rho^{(1)}_{12}}{1-\rho^{(1)}_{12}}\Big]\nonumber\\
&-&{\rm{Li}}_{2}\Big[1-\frac{\rho^{(+)}_{1}}{1-\alpha-\rho^{(+)}_{1}}
\frac{\rho^{(2)}_{12}}{1-\rho^{(2)}_{12}}\Big]
-{\rm{Li}}_{2}\Big[\frac{1-\alpha}{\rho^{(-)}_{1}}\Big]\nonumber\\
&+&{\rm{Li}}_{2}\Big[1-\frac{\rho^{(-)}_{1}}{1-\alpha-\rho^{(-)}_{1}}
\frac{\rho^{(1)}_{12}}{1-\rho^{(1)}_{12}}\Big]
+{\rm{Li}}_{2}\Big[1-\frac{\rho^{(-)}_{1}}{1-\alpha-\rho^{(-)}_{1}}
\frac{\rho^{(2)}_{12}}{1-\rho^{(2)}_{12}}\Big]\Big\}.   \label{finalresultofintIonetwo}
\end{eqnarray}

\begin{itemize}
\item evaluation of $I_{13}$
\end{itemize}
The integral is
\begin{equation}
I_{13}=-\alpha\int_{0}^{1}\D \rho\,
\frac{1}{F_{13}(\rho)}
\Big\{\ln\Big[(g_{0}-D_{1} \alpha)\rho\Big]
-\ln G_{13}(\rho)\Big\},   \label{newformofintIonethree}
\end{equation}
where
\begin{eqnarray}
F_{13}(\rho)&=&\alpha^{2}(A_{2}g_{0}-A_{1}D_{1})\rho^{2}
+\alpha(B_{1}D_{1}-B_{2}g_{0})\rho
-g_{0}(\mu_{s}^{2}+i\epsilon)\nonumber\\
&=&a_{13}[\rho-\rho^{(+)}_{13}][\rho-\rho^{(-)}_{13}],   \label{functionFonethree}
\end{eqnarray}
with
\begin{equation}
a_{13}=\alpha^{2}(A_{2}g_{0}-A_{1}D_{1}),\quad\quad
\rho^{(+)}_{13}=-\frac{\rho^{(+)}_{1}}{\alpha},\quad\quad
\rho^{(-)}_{13}=-\frac{\rho^{(-)}_{1}}{\alpha},  \label{twozerosoffunconethree}
\end{equation}
The argument of second logarithm in the numerator is
\begin{eqnarray}
G_{13}(u)&=&\alpha  (\alpha A_{2} -A_{1})\rho^{2}
-(\alpha B_{2}-B_{1} )\rho-
\mu_{s}^{2}-i\epsilon\nonumber\\
&=&\frac{G_{13}(1)[\rho-\rho^{(1)}_{13}][\rho-\rho^{(2)}_{13}]}
{[1-\rho^{(1)}_{13}][1-\rho^{(2)}_{13}]},  \label{functionGonethree}
\end{eqnarray}
with
\begin{eqnarray}
G_{13}(1)&=&m_{2}^{2}-\mu_{s}^{2}-i\epsilon\nonumber\\
\rho^{(1)}_{13}&=&\frac{1}{2\alpha(\alpha A_{2}-A_{1})}
\Big[-(B_{1}-\alpha B_{2})+\sqrt{\Delta_{13}}\Big]\nonumber\\
\rho^{(2)}_{13}&=&\frac{1}{2\alpha(\alpha A_{2}-A_{1})}
\Big[-(B_{1}-\alpha B_{2})-\sqrt{\Delta_{13}}\Big]\nonumber\\
\Delta_{13}&=&(B_{1}-\alpha B_{2})^{2}
+4\alpha(\alpha A_{2}-A_{1})(\mu_{s}^{2}+i\epsilon),  \label{zerosoffuncGonethree}
\end{eqnarray}
where Eq.(\ref{constrainofintegrationone}) has been employed.
By using the equations listed in appendix B, it is easily to obtain
\begin{eqnarray}
I_{13}&=&-\alpha\int_{0}^{1}\D \rho\,
\frac{1}{a_{13}[\rho-\rho^{(+)}_{13}][\rho-\rho^{(+)}_{13}]}
\Big\{\ln\Big[(g_{0}-D_{1} \alpha)\rho\Big]
-\ln\frac{G_{13}(1)\big[\rho-\rho^{(1)}_{13}\big]\big[\rho-\rho^{(2)}_{13}\big]}
{\big[1-\rho^{(1)}_{13}\big]\big[1-\rho^{(2)}_{13}\big]}\Big\}\nonumber\\
&=&\frac{-\alpha}{a_{13}\big[\rho^{(+)}_{13}-\rho^{(-)}_{13}\big]}
\int_{0}^{1}\D \rho\,
\Big[\frac{1}{\rho-\rho^{(+)}_{13}}
-\frac{1}{\rho-\rho^{(-)}_{13}}\Big]\nonumber\\
&\times&\Big\{\ln\Big[(g_{0}-D_{1} \alpha)\rho\Big]
-\ln\frac{G_{13}(1)}{\mu^{2}}
-\ln\frac{[\rho-\rho^{(1)}_{13}][\rho-\rho^{(2)}_{13}]}
{[1-\rho^{(1)}_{13}][1-\rho^{(2)}_{13}]}\,\Big\}\nonumber\\
&=&\frac{1}{a_{1}\big[\rho^{(+)}_{1}-\rho^{(-)}_{1}\big]}
\Big\{\frac{1}{2}\ln^{2}\frac{\lambda^{2}}{\mu_{s}^{2}}
+\ln\alpha\ln\frac{\lambda^{2}}{\mu_{s}^{2}}
+\frac{1}{2}\ln^{2}\alpha
+\ln\Big(1+\frac{\alpha\lambda^{2}}{\mu_{s}^{2}}\Big)
\ln\frac{g_{0}-\alpha D_{1}}{G_{13}(1)}\nonumber\\
&+&\frac{\pi^{2}}{6}
-\ln\Big[1+\frac{\alpha}{\rho^{(-)}_{1}}\Big]
\ln\frac{g_{0}-\alpha D_{1}}{G_{13}(1)}\nonumber\\
&+&{\rm{Li}}_{2}\Big(-\frac{\mu_{s}^{2}}{\alpha\lambda^{2}}\Big)
+{\rm{Li}}_{2}\Big[1+\frac{\rho^{(+)}_{1}}{\alpha+\rho^{(+)}_{1}}
\frac{\rho^{(1)}_{13}}{1-\rho^{(1)}_{13}}\Big]
+{\rm{Li}}_{2}\Big[1+\frac{\rho^{(+)}_{1}}{\alpha+\rho^{(+)}_{1}}
\frac{\rho^{(2)}_{13}}{1-\rho^{(2)}_{13}}\Big]\nonumber\\
&+&{\rm{Li}}_{2}\Big[-\frac{\alpha}{\rho^{(-)}_{1}}\Big]
-{\rm{Li}}_{2}\Big[1+\frac{\rho^{(-)}_{1}}{\alpha+\rho^{(-)}_{1}}
\frac{\rho^{(1)}_{13}}{1-\rho^{(1)}_{13}}\Big]
-{\rm{Li}}_{2}\Big[1+\frac{\rho^{(-)}_{1}}{\alpha+\rho^{(-)}_{1}}
\frac{\rho^{(2)}_{13}}{1-\rho^{(2)}_{13}}\Big]\Big\}.   \label{finalresultofIonethree}
\end{eqnarray}

\section{evaluating the three components of $I_{0}$}
\begin{itemize}
\item  evaluation of $I_{01}$
\end{itemize}
The integral is
\begin{eqnarray}
I_{01}&=&\int_{0}^{1}\D \rho\,\frac{1}{g_{0}h_{2}-g_{1}h_{1}}
\Big[\ln(g_{0}+g_{1})-\ln(h_{1}+h_{2})\Big]\nonumber\\
&=&\int_{0}^{1}\D \rho\,\frac{1}{F_{01}(\rho)}
\Big[\ln\Big(D_{1}\rho+g_{0}\Big)
-\ln G_{01}(\rho)\Big],   \label{intIzeroonenew}
\end{eqnarray}
where the denominator is
\begin{eqnarray}
F_{01}(u)&=&(K_{2}g_{0}-K_{1}D_{1})\rho^{2}
+(g_{0}N_{2}-N_{1}D_{1})\rho-g_{0}(\mu_{s}^{2}+i\epsilon)\nonumber\\
&=&a_{01}\big[\rho-\rho^{(+)}_{01}\big]\big[\rho-\rho^{(-)}_{01}\big],  \label{factorizationoffunczeroone}
\end{eqnarray}
with
\begin{eqnarray}
a_{01}=a_{0}&=&K_{2}g_{0}-K_{1}D_{1}\nonumber\\
\rho^{(+)}_{01}=\rho^{(+)}_{0}&=&
\frac{\lambda_{0}^{2}}{\mu_{s}^{2}}+i\epsilon,\quad\quad
\rho^{(-)}_{01}=\rho^{(-)}_{0}
=-(\kappa_{0}+\frac{\lambda_{0}^{2}}{\mu_{s}^{2}})-i\epsilon,  \label{twozerosoffunczeroone}
\end{eqnarray}
where in order to simplifying the symbols, we have
relabel $a_{01}$, $\rho^{(+)}_{01}$ and $\rho^{(-)}_{01}$ as
$a_{0}$, $\rho^{(+)}_{0}$ and $\rho^{(-)}_{0}$, respectively.
The dimensional $\lambda_{0}$ and dimensionless $\kappa_{0}$ are
defined as follows
\begin{equation}
\lambda_{0}^{2}=\frac{N_{2}g_{0}-N_{1}D_{1}}{g_{0}},\quad\quad
\kappa_{0}=\frac{N_{2}g_{0}-N_{1}D_{1}}{K_{2}g_{0}-K_{1}D_{1}},  \label{lambdaandkappazero}
\end{equation}
The argument of the second logarithm in numerator is
\begin{eqnarray}
G_{01}(\rho)&=&K_{2}\rho^{2}+(K_{1}+N_{2})\rho+N_{1}-\mu_{s}^{2}-i\epsilon\nonumber\\
&=&\frac{G_{01}(1)\big[\rho-\rho^{(1)}_{01}\big]\big[\rho-\rho^{(2)}_{01}\big]}
{\big[1-\rho^{(1)}_{01}\big]\big[1-\rho^{(2)}_{01}\big]},  \label{functGonezerofactorized}
\end{eqnarray}
with
\begin{eqnarray}
G_{01}(1)&=&m_{3}^{2}-s_{12}^{2}
+\beta(m_{3}^{2}+2\omega_{3}^{2}+s_{12}^{2}-m_{2}^{2}-m_{4}^{2})
-\mu_{s}^{2}-i\epsilon \nonumber\\
\rho^{(1)}_{01}&=&\frac{1}{2K_{2}}[-(K_{1}+N_{2})+\sqrt{\Delta_{01}}]\nonumber\\
\rho^{(2)}_{01}&=&\frac{1}{2K_{2}}[-(K_{1}+N_{2})-\sqrt{\Delta_{01}}]\nonumber\\
\Delta_{01}&=&(K_{1}+N_{2})^{2}-4K_{2}(N_{1}-\mu_{s}^{2}-i\epsilon),  \label{zerosoffunctGzeoone}
\end{eqnarray}
With the help of equations listed in appendix B, we find
\begin{eqnarray}
I_{01}&=&\int_{0}^{1}\D \rho\,
\frac{1}{a_{0}\big[\rho-\rho^{(+)}_{0}\big]\big[\rho-\rho^{(-)}_{0}\big]}
\Big[\ln\Big(D_{1}\rho+g_{0}\Big)-\ln G_{01}(\rho)\Big]\nonumber\\
&=&\frac{1}{a_{0}\big[\rho^{(+)}_{0}-\rho^{(-)}_{0}\big]
}\int_{0}^{1}\D \rho\,
\Big[\frac{1}{\rho-\rho^{(+)}_{0}}
-\frac{1}{\rho-\rho^{(-)}_{0}}\Big]\nonumber\\
&\times&\Big\{\ln\Big(D_{1}\rho+g_{0}\Big)
-\ln \frac{G_{01}(1)}{\mu^{2}}
-\ln\frac{\big[\rho-\rho^{(1)}_{01}\big]\big[\rho-u^{(2)}_{01}\big]}
{\big[1-\rho^{(1)}_{01}\big]\big[1-\rho^{(2)}_{01}\big]}\,\Big\}\nonumber\\
&=&\frac{1}{a_{0}\big[\rho^{(+)}_{0}-\rho^{(-)}_{0}\big]}
\Big\{\ln^{2}\frac{\lambda_{0}^{2}}{\mu_{s}^{2}}
+\ln\Big|1-\frac{\lambda_{0}^{2}}{\mu_{s}^{2}}\Big|
\ln\frac{g_{0}+D_{1}\rho^{(+)}_{0}}{G_{01}(1)}
+2i\pi\ln\frac{\lambda_{0}^{2}}{\mu_{s}^{2}}\nonumber\\
&+&i\pi\ln\frac{g_{0}+D_{1}\rho^{(+)}_{0}}{G_{01}(1)}
-\frac{2\pi^{2}}{3}
-\ln\frac{g_{0}+D_{1}\rho^{(-)}_{0}}{G_{01}(1)}\ln\Big[1-\frac{1}{\rho^{(-)}_{0}}\Big]\nonumber\\
&+&{\rm{Li}}_{2}\Big[\frac{-D_{1}\rho^{(+)}_{0}}{-D_{1}\rho^{(+)}_{0}-g_{0}}\Big]
-{\rm{Li}}_{2}\Big[\frac{D_{1}(1-\rho^{(+)}_{0})}{-D_{1}\rho^{(+)}_{0}-g_{0}}\Big]\nonumber\\
&-&{\rm{Li}}_{2}\Big[\frac{-D_{1}\rho^{(-)}_{0}}{-D_{1}\rho^{(-)}_{0}-g_{0}}\Big]
+{\rm{Li}}_{2}\Big[\frac{D_{1}(1-\rho^{(-)}_{0})}{-D_{1}\rho^{(-)}_{0}-g_{0}}\Big]\nonumber\\
&+&2{\rm{Li}}_{2}\Big(\frac{\mu_{s}^{2}}{\lambda_{0}^{2}}\Big)
+{\rm{Li}}_{2}\Big[1-\frac{\rho^{(+)}_{0}}{1-\rho^{(+)}_{0}}
\frac{\rho^{(1)}_{01}}{1-\rho^{(1)}_{01}}\Big]
+{\rm{Li}}_{2}\Big[1-\frac{\rho^{(+)}_{0}}{1-\rho^{(+)}_{0}}
\frac{\rho^{(2)}_{01}}{1-\rho^{(2)}_{01}}\Big]\nonumber\\
&+&2{\rm{Li}}_{2}\Big[\frac{1}{\rho^{(-)}_{0}}\Big]
-{\rm{Li}}_{2}\Big[1-\frac{\rho^{(-)}_{0}}{1-\rho^{(-)}_{0}}
\frac{\rho^{(1)}_{01}}{1-\rho^{(1)}_{01}}\Big]
-{\rm{Li}}_{2}\Big[1-\frac{\rho^{(-)}_{0}}{1-\rho^{(-)}_{0}}
\frac{\rho^{(2)}_{01}}{1-\rho^{(2)}_{01}}\Big]\Big\}.  \label{finalresultsofIzeroone}
\end{eqnarray}

\begin{itemize}
\item  evaluation of $I_{02}$
\end{itemize}

The integral is
\begin{equation}
I_{02}=(1-\beta)\int_{0}^{1}\D
\rho\,\frac{1}{F_{02}(\rho)}
\Big\{-\ln\Big[\Big(g_{0}+(1-\beta)D_{1}\Big)\rho\Big]
+\ln G_{02}(\rho)\Big\},   \label{newformofintIzerotwo}
\end{equation}
where the denominator is
\begin{eqnarray}
F_{02}(\rho)&=&(1-\beta)^{2}(K_{2}g_{0}-K_{1}D_{1})\rho^{2}
-(1-\beta)(N_{1}D_{1}-N_{2}g_{0})\rho-
g_{0}(\mu_{s}^{2}+i\epsilon)\nonumber\\
&=&a_{02}\big[\rho-\rho^{(+)}_{02}\big]
\big[\rho-\rho^{(-)}_{02}\big],  \label{functionFzerotwo}
\end{eqnarray}
with
\begin{equation}
a_{02}=(1-\beta)^{2}(K_{2}g_{0}-K_{1}D_{1}),\quad\quad
\rho^{(+)}_{02}=\frac{\rho^{(+)}_{0}}{1-\beta},\quad\quad
\rho^{(-)}_{02}=\frac{\rho^{(-)}_{0}}{1-\beta},  \label{twozerosoffunczeotwo}
\end{equation}
The argument of the second logarithm in the numerator is
\begin{eqnarray}
G_{02}(\rho)&=&(1-\beta )[K_{1}+(1- \beta )K_{2}]\rho^{2}
+[N_{1}+(1- \beta)N_{2}]\rho-\mu_{s}^{2}-i\epsilon\nonumber\\
&=&\frac{G_{02}(1)\big[\rho-\rho^{(1)}_{02}\big]
\big[\rho-\rho^{(2)}_{02}\big]}
{\big[1-\rho^{(1)}_{02}\big]\big[1-\rho^{(2)}_{02}\big]},  \label{functGzerotwofactorized}
\end{eqnarray}
with
\begin{eqnarray}
G_{02}(1)&=&m_{3}^{2}-\mu_{s}^{2}-i\epsilon\nonumber\nonumber\\
\rho^{(1)}_{02}&=&\frac{1}{2(1-\beta)[K_{1}+(1-\beta)K_{2}]}
\Big\{-[N_{1}+(1-\beta)N_{2}]+\sqrt{\Delta_{02}}\Big\}\nonumber\\
\rho^{(2)}_{02}&=&\frac{1}{2(1-\beta)[K_{1}+(1-\beta)K_{2}]}
\Big\{-[N_{1}+(1-\beta)N_{2}]-\sqrt{\Delta_{02}}\Big\}\nonumber\\
\Delta_{02}&=&\Big[N_{1}+(1-\beta)N_{2}]^{2}
+4(1-\beta)[K_{1}+(1-\beta)K_{2}](\mu_{s}^{2}+i\epsilon),  \label{zerosoffunctGzerotwo}
\end{eqnarray}
where Eq.(\ref{constrainofbeta}) has been employed.
By using the equations in the appendix B, it is not difficult to
get the result
\begin{eqnarray}
I_{02}&=&(1-\beta)\int_{0}^{1}\D \rho\,
\frac{1}{a_{02}\big[\rho-\rho^{(+)}_{02}\big]
\big[\rho-\rho^{(-)}_{02}\big]}
\Big\{-\ln\Big[\Big(g_{0}+(1-\beta)D_{1}\Big)\rho\Big]
+\ln G_{02}(\rho)\Big\}\nonumber\\
&=&\frac{1-\beta}{a_{02}\big[\rho^{(+)}_{02}-\rho^{(-)}_{02}\big]}
\int_{0}^{1}\D \rho\,
\Big[\frac{1}{\rho-\rho^{(+)}_{02}}-\frac{1}{\rho-\rho^{(-)}_{02}}\Big]
\Big\{-\ln\Big[\Big(g_{0}+(1-\beta)D_{1}\Big)\rho\Big]\nonumber\\
&+&\ln\frac{G_{02}(1)}{\mu^{2}}
+\ln\frac{\big[\rho-\rho^{(1)}_{02}\big]\big[\rho-\rho^{(2)}_{02}\big]}
{\big[1-\rho^{(1)}_{02}\big]\big[1-\rho^{(2)}_{02}\big]}\,\Big\}\nonumber\\
&=& \frac{1}{a_{0}\big[\rho^{(+)}_{0}-\rho^{(-)}_{0}\big]}
\Big\{-\frac{1}{2}\ln^{2}\frac{\lambda_{0}^{2}}{\mu_{s}^{2}}
-\ln(1-\beta)\ln\frac{\lambda_{0}^{2}}{\mu_{s}^{2}}
-\frac{1}{2}\ln^{2}(1-\beta)\nonumber\\
&-&\ln\Big|1-\frac{(1-\beta)\lambda_{0}^{2}}{\mu_{s}^{2}}\Big|
\ln\frac{g_{0}+(1-\beta)D_{1}}{G_{02}(1)}
-i\pi\ln\frac{\lambda_{0}^{2}}{\mu_{s}^{2}}
-i\pi\ln(1-\beta)\nonumber\\
&-&i\pi\ln\frac{g_{0}+(1-\beta)D_{1}}{G_{02}(1)}
+\frac{\pi^{2}}{3}
+\ln\frac{g_{0}+(1-\beta)D_{1}}{G_{02}(1)}
\ln\Big[1-\frac{1-\beta}{\rho^{(-)}_{0}}\Big]\nonumber\\
&-&{\rm{Li}}_{2}\Big[\frac{\mu_{s}^{2}}{(1-\beta)\lambda_{0}^{2}}\Big]
-{\rm{Li}}_{2}\Big[1-\frac{\rho^{(+)}_{0}}{1-\beta-\rho^{(+)}_{0}}
\frac{\rho^{(1)}_{02}}{1-\rho^{(1)}_{02}}\Big]
-{\rm{Li}}_{2}\Big[1-\frac{\rho^{(+)}_{0}}{1-\beta-\rho^{(+)}_{0}}
\frac{\rho^{(2)}_{02}}{1-\rho^{(2)}_{02}}\Big]\nonumber\\
&-&{\rm{Li}}_{2}\Big[\frac{1-\beta}{\rho^{(-)}_{0}}\Big]
+{\rm{Li}}_{2}\Big[1-\frac{\rho^{(-)}_{0}}{1-\beta-\rho^{(-)}_{0}}
\frac{\rho^{(1)}_{02}}{1-\rho^{(1)}_{02}}\Big]
+{\rm{Li}}_{2}\Big[1-\frac{\rho^{(-)}_{0}}{1-\beta-\rho^{(-)}_{0}}
\frac{\rho^{(2)}_{02}}{1-\rho^{(2)}_{02}}\Big]
\Big\}.                        \label{resultsofintIzerotwo}
\end{eqnarray}

\begin{itemize}
\item  evaluation of $I_{03}$
\end{itemize}
The integral is
\begin{eqnarray}
I_{03}&=&-\beta\int_{0}^{1}\D \rho\,
\frac{1}{F_{03}(\rho)}\Big\{\ln\Big[\Big(g_{0}-\beta D_{1} \Big)\rho\Big]
-\ln G_{03}(\rho)\Big\},  \label{newformofintIzerothree}
\end{eqnarray}
where we define
\begin{eqnarray}
F_{03}(\rho)&=&\beta^{2}(K_{2}g_{0}-K_{1}D_{1})\rho^{2}
+\beta(N_{1}D_{1}-N_{2}g_{0})\rho-g_{0}(\mu_{s}^{2}+i\epsilon)\nonumber\\
&=&a_{03}\big[\rho-\rho^{(+)}_{03}\big]\big[\rho-\rho^{(-)}_{03}\big],   \label{factorizationoffunctionFzerothr}
\end{eqnarray}
with
\begin{equation}
a_{03}=\beta^{2}(K_{2}g_{0}-K_{1}D_{1}),\quad\quad
\rho^{(+)}_{03}=-\frac{\rho^{(+)}_{0}}{\beta},
\quad\quad \rho^{(-)}_{03}=-\frac{\rho^{(-)}_{0}}{\beta},   \label{zerosoffunzerothr}
\end{equation}
The argument of the second logarithm in the numerator is
\begin{eqnarray}
G_{03}(\rho)&=&\beta  (\beta K_{2}-K_{1})\rho^{2}
-(\beta N_{2} -N_{1})\rho-\mu_{s}^{2}-i\epsilon\nonumber\\
&=&\frac{G_{03}(1)\big[\rho-\rho^{(1)}_{03}\big]\big[\rho-\rho^{(2)}_{03}\big]}
{\big[1-\rho^{(1)}_{03}\big]\big[1-\rho^{(2)}_{03}\big]}, \label{factorizationofGzerothr}
\end{eqnarray}
with
\begin{eqnarray}
G_{03}(1)&=&m_{2}^{2}-\mu_{s}^{2}-i\epsilon\nonumber\\
\rho^{(1)}_{03}&=&\frac{1}{2\beta(\beta K_{2}-K_{1})}[-(N_{1}-\beta N_{2})+\sqrt{\Delta_{03}}]\nonumber\\
\rho^{(2)}_{03}&=&\frac{1}{2\beta(\beta K_{2}-K_{1})}[-(N_{1}-\beta N_{2})-\sqrt{\Delta_{03}}]\nonumber\\
\Delta_{03}&=&(N_{1}-\beta N_{2})^{2}+4\beta(\beta K_{2}-K_{1})(\mu_{s}^{2}+i\epsilon),  \label{zerosoffunctGzerothr}
\end{eqnarray}
where Eq.(\ref{constrainofbeta}) has been used.
By employing the equations in the appendix B we can get
\begin{eqnarray}
I_{03}&=&-\beta\int_{0}^{1}\D \rho\,
\frac{1}{a_{03}\big[\rho-\rho^{(+)}_{03}\big]\big[\rho-\rho^{(-)}_{03}\big]}
\Big\{\ln\Big[\Big(g_{0}-\beta D_{1} \Big)\rho\Big]
-\ln G_{03}(\rho)\Big\}\nonumber\\
&=&\frac{-\beta}{a_{03}\big[\rho^{(+)}_{03}-\rho^{(-)}_{03}\big]}
\int_{0}^{1}\D \rho\,\Big[\frac{1}{\rho-\rho^{(+)}_{03}}-\frac{1}{\rho-\rho^{(-)}_{03}}\Big]
\Big\{\ln\Big[\Big(g_{0}-\beta D_{1} \Big)\rho\Big]\nonumber\\
&-&\ln\frac{G_{03}(1)}{\mu^{2}}
-\ln\frac{\big[\rho-\rho^{(1)}_{03}\big]\big[\rho-\rho^{(1)}_{03}\big]}
{\big[1-\rho^{(1)}_{03}\big]\big[1-\rho^{(2)}_{03}\big]}\Big\}\nonumber\\
&=&\frac{1}{a_{0}\big[\rho^{(+)}_{0}-\rho^{(-)}_{0}\big]}
\Big\{\frac{1}{2}\ln^{2}\frac{\lambda_{0}^{2}}{\mu_{s}^{2}}
+\ln\beta\ln\frac{\lambda_{0}^{2}}{\mu_{s}^{2}}
+\frac{1}{2}\ln^{2}\beta
+\ln\Big|1+\frac{\beta\lambda_{0}^{2}}{\mu_{s}^{2}}\Big|
\ln\frac{g_{0}-\beta D_{1}}{G_{03}(1)}\nonumber\\
&+&\frac{\pi^{2}}{6}
-\ln\Big[1+\frac{\beta}{\rho^{(-)}_{0}}\Big]
\ln\frac{g_{0}-\beta D_{1}}{G_{03}(1)}
+{\rm{Li}}_{2}\Big(-\frac{\mu_{s}^{2}}{\beta\lambda_{0}^{2}}\Big)\nonumber\\
&+&{\rm{Li}}_{2}\Big[1+\frac{\rho^{(+)}_{0}}{\beta+\rho^{(+)}_{0}}\
\frac{\rho^{(1)}_{03}}{1-\rho^{(1)}_{03}}\Big]
+{\rm{Li}}_{2}\Big[1+\frac{\rho^{(+)}_{0}}{\beta+\rho^{(+)}_{0}}
\frac{\rho^{(2)}_{03}}{1-\rho^{(2)}_{03}}\Big]\nonumber\\
&+&{\rm{Li}}_{2}\Big[-\frac{\beta}{\rho^{(-)}_{0}}\Big]
-{\rm{Li}}_{2}\Big[1+\frac{\rho^{(-)}_{0}}{\beta+\rho^{(-)}_{0}}
\frac{\rho^{(1)}_{03}}{1-\rho^{(1)}_{03}}\Big]
-{\rm{Li}}_{2}\Big[1+\frac{\rho^{(-)}_{0}}{\beta+\rho^{(-)}_{0}}
\frac{\rho^{(2)}_{03}}{1-\rho^{(2)}_{03}}\Big]\Big\}.  \label{finalresultsoofIzerothr}
\end{eqnarray}

\end{document}